\documentclass[aps,prx,twocolumn,10pt,superscriptaddress,longbibliography]{revtex4-2}

\usepackage[utf8]{inputenc}     
\usepackage{amsmath,amssymb}     
\usepackage{braket}
\usepackage{graphicx}             
\usepackage{xcolor}              
\usepackage[colorlinks=true, linkcolor=blue, citecolor=blue, urlcolor=blue]{hyperref}
\usepackage[capitalise]{cleveref}
\usepackage{ragged2e}
\usepackage[caption=false]{subfig}
\usepackage{comment}

\begin{document}

\title{Trapped-Ion Multiqubit Gates are Compatible with Scalable Quantum Error Correction}
\author{Ori Grossman}
\affiliation{Quantum Art Ltd, Ness Ziona 7403682, Israel}

\email{ori.grossman@quantum-art.tech}

\author{Yotam Kadish}
\affiliation{Quantum Art Ltd, Ness Ziona 7403682, Israel}

\author{Snir Gazit}
\affiliation{Quantum Art Ltd, Ness Ziona 7403682, Israel}
\author{Amit Ben-Kish}
\affiliation{Quantum Art Ltd, Ness Ziona 7403682, Israel}

\author{Roee Ozeri}
\affiliation{Quantum Art Ltd, Ness Ziona 7403682, Israel}
\affiliation{Department of Physics of Complex Systems, Weizmann Institute of Science, Rehovot 7610001, Israel}

\author{Yotam Shapira}
\affiliation{Quantum Art Ltd, Ness Ziona 7403682, Israel}

\date{\today}
\begin{abstract}

We construct a detailed microscopic noise model for multi-qubit (MQ) gate operations in the context of trapped ion architecture with all-to-all connectivity.
We find that phonon heating and motional dephasing are well captured by effective single- and two-qubit error channels that can, in principle, act between arbitrary pairs of qubits. Nevertheless, the median magnitude of two-qubit errors between uncoupled qubits is substantially smaller than that of errors between gate-coupled qubits. Errors associated with photon scattering are shown to solely propagate to qubits participating in gate operations. Lastly, we combine all noise sources, assigned with experimentally relevant parameters, and explore the scalability of a quantum error correction (QEC) scheme based on the rotated surface code, as a function of error rates and code size.
Our analysis bridges device-level physics and QEC performance for MQ gates in trapped-ion architectures.
\end{abstract}

\maketitle

\section{Introduction}
\label{section:intro}
Quantum computing is rapidly advancing, with recent progress marking important steps toward quantum advantage \cite{gharibyan2025heuristic,kim2023evidence,google2025observation}. The fragile nature of quantum information in the presence of various noise sources remains a major obstacle, leading to prohibitively high error rates that severely limit the depth of computation \cite{niroula2025realizationquantumstreamingalgorithm,he2025performance}. In this regard, quantum error correction (QEC) offers a systematic approach to overcome these limitations via efficient schemes for encoding and decoding of logical qubits into multiple physical qubits \cite{aharonov1999faulttolerantquantumcomputationconstant,shor1995scheme,steane1996error}. The last decade brought significant advances in demonstrating the fundamental building blocks of QEC across various quantum computing hardware \cite{Zobrist2024, sales2025experimental, dasu2025breaking}.

A leading modality for quantum computing is trapped ions, as they enable high-fidelity operations \cite{hughes2025trapped,smith2025single}, owing to their remarkably long coherence times and repeatability. The long-range Coulomb interactions among the ions allow for near-arbitrary all-to-all connectivity. This unique property enables multiqubit (MQ) gate operations through engineered spin–motion interactions. Indeed, several recent works have demonstrated programmable, simultaneous all-to-all entangling gates \cite{grzesiak2020efficient,shapira2023fastdesignscalingmultiqubit}. This mode of operation should be contrasted with typical experimental implementations, which rely on a combination of two-qubit operations with mechanical charge shuttling \cite{ransford2025helios}. MQ gates have been shown to improve performance and reduce circuit depth in various applications, including optimization problems, the quantum Fourier transform, quantum volume circuits, Clifford circuits, N-qubit Toffoli gates, and a wide range of random unitaries \cite{Schwerdt2022, grzesiak2020efficient, Bravyi2022compilation, Nemirovsky2025efficient_compilation, Baler2023, Hyer2005fanout, Foxman2025PRUs, Decross2025random_unitaries}.

In the context of QEC, long-range connectivity enables the measurement of nonlocal stabilizers, as required by quantum LDPC and related codes \cite{Panteleev2021, Wang2026}. 
Such connectivity also facilitates parallel extraction of syndromes, including for stabilizers that overlap on common data qubits, independently of their weights. In addition, any logical product state of a CSS code can be prepared using a single MQ gate~\cite{sheffer2024preparingtopologicalstatesfinite} (though not fault-tolerantly). Lastly, we note that there has been significant progress in implementing Clifford gates using all-to-all connectivity with a constant number of MQ gates~\cite{nemirovsky2025optimalconstantcostimplementationsclifford}.

Despite the advantages offered by MQ gates, it is a priori unclear whether they outweigh the potential long-range error propagation and ultimately permit fault-tolerant quantum computation. Devising noise models for such systems requires great care. Unlike two-qubit gate-based architectures, where the native gates are relatively simple and the associated noise channels are relatively well understood \cite{Ballance2016, sutherland2022one,lotshaw2023modeling}, for MQ gates, the situation is more challenging. The effect of noise, arising from the interplay between the collective modes of motion and a large number of qubits, is highly non-trivial. MQ gates involving $N\gg2$ qubits require a large set of Pauli operators for a complete and accurate description of the noise channel (in general $4^N$).  
It is therefore crucial to develop a compact yet reliable description of these noise channels in order to properly assess MQ gate performance, and tailor suitable codes and decoders for QEC.

Several physically motivated noise models for trapped ion systems have been proposed based on the dominant experimental noise sources \cite{PhysRevX.11.041058, Trout2018,Liu2025,16gx-m7kg}. However, these models are primarily tailored to two-qubit gates and are not directly applicable to MQ gates, where a more careful microscopic treatment is required. In this work, we construct a detailed noise model for MQ gates. A common classification of error sources \cite{ozeri2007errors} is: classical noise and quantum noise. The origin of classical noise can be divided into gate imperfection errors and system noise. Imperfection errors are coherent in nature and stem from a variety of miscalibrations during gate operation, such as the laser power and mode characterization, among others. Coherent control methods are typically applied to tame their effect \cite{shapira2018robust,shapira2023fastdesignscalingmultiqubit,shapira2020theory,shapira2023robust,kang2021batch,milne2020phase,leung2018robust}. Such errors are, quite generically, simpler to analyze, as they give rise to single- or two-qubit overrotations. We note that, to treat them as stochastic Pauli channels for QEC, Pauli twirling is required. These error sources, in the MQ context, are discussed in Ref.~\cite{Liu2025}. 
System noise sources are stochastic in nature, and their overall effect on qubits is less straightforward. Typical noise sources of this kind include phonon-mode dephasing or heating.

Quantum noise, on the other hand, is a fundamental source of error that cannot be completely mitigated. A case in point is Photon scattering, which is intrinsically tied to transitions utilized for qubit operations \cite{ozeri2007errors,uys2010decoherence}. Since the probability of this event is set by coupling to the electromagnetic vacuum, it is an unavoidable source of error, and as such it is particularly important to study. Raman scattering, together with other stochastic noise sources, were studied numerically for a specific gate operation in Ref.~\cite{Schwerdt2022}, but an overarching analytic picture that can provide insights for MQ-gate-based QEC is lacking. 

Here, we mainly consider the effect of scattering events, which, without loss of generality, can be treated as depolarizing noise. Similarly effective qubit dephasing can arise due to magnetic field fluctuation, especially if the qubit levels are first order sensitive to the magnetic field (e.g. qubits based on Zeeman splitting). In addition, we investigate the stochastic noise channel associated with  the effect of noise in the crystal normal-mode frequencies leading to motional heating and dephasing, which previous works have shown to result in a sizable contribution to two-qubit gate infidelity \cite{Orozco-Ruiz2025,Ballance2016}.

Our key finding is that photon-scattering errors lead to correlated errors exclusively between qubits coupled to the qubit on which photon scattering occurred via the MQ gate. Therefore, noise propagation follows the connectivity structure imposed by the MQ gate operation. For motional modes, we find that mode heating during the evolution of the MQ gate is captured by a single-qubit error, while motional dephasing generates both single and two-qubit errors to leading order, similarly to the effects of coherent errors. Notably, two-qubit errors are primarily restricted to qubits that are coupled at gate time.
A summary of these results is shown in Table \ref{table1}.
Furthermore, we demonstrate that the rotated surface code, executed using MQ gates, shows a finite-threshold behavior when two-qubit errors between uncoupled qubits are negligible. Even when correlated two-qubit errors between uncoupled qubits become appreciable, a practical crossing point corresponding to gains for small code distances still persists. Nevertheless, for a well-defined finite threshold to exist, the total contribution of two-qubit errors per qubit must remain intensive rather than extensive~\cite{Fowler2014Nonlocal,Aharonov2006LongRange}.

\begin{table}[h] 
    \centering
    \begin{tabular}{|c|c|c|}
        \hline
        Noise & Type & Pauli weight \\ \hline
        Miscalibration & Coherent& 1q, 2q~\cite{Liu2025}  \\ \hline
        Phonon heating & Classical & 1q \\ \hline
        Phonon dephasing & Classical & 1q, 2q \\ \hline
        Spin dephasing & Classical & 1q, MQ-connected\\ \hline
        Spin depolarization &  Quantum & 1q, MQ-connected \\ \hline
    \end{tabular}    
    \caption{Summary of the different noise sources impacting multiqubit gates and their effect. Below we use these results to model the noise channels, which in turn are delivered to a QEC protocol such as the surface code.
    }
\label{table1}
\end{table}

The remainder of the paper is organized as follows: We begin by reviewing trapped ions Hamiltonian dynamics, governing MQ gates, in \cref{section:MQ_hamiltonian}. In \cref{section:noise_sources}, we examine the various noise sources and associated jump operators within the Lindblad formalism. Next, in \cref{section:analytical}, we present a detailed analytical and numerical modeling for the various noise channels, including scattering noise (or spin dephasing), heating, and motional dephasing. Lastly, in \cref{section:QEC} we combine all noise sources and demonstrate a robust QEC within the rotated surface code utilizing MQ gates. We conclude in \cref{section:discussion} by summarizing our results and discussing open research directions for QEC based on MQ gates and potential implementations.

\section{Multiqubit gate model}
\label{section:MQ_hamiltonian}
In trapped-ion platforms, qubits are typically encoded in the  electronic (“internal”) states of the ions, forming effective two-level systems. Entangling gates between qubits are realized by electromagnetically coupling these internal states to vibrational (phonon) modes. This phonon-mediated interaction underlies high-fidelity entangling gates \cite{Ballance2016,hughes2025trapped}. A prominent example is the celebrated Mølmer–Sørensen (MS) gate~\cite{Srensen1999}, which utilizes a single center-of-mass (COM) mode.

In this work, we focus on generalizations of the MS gate that non-trivially take advantage of multiple phonon modes. This enables the implementation of generalized all-to-all entangling Ising gates of the form,
\begin{equation}
U_{\mathrm{Ising}}=e^{\,i \sum_{n<m}\varphi_{nm}{\sigma}^{(n)}_x {\sigma}^{(m)}_x} .
\label{eq:Umq}
\end{equation}
Here, ${\sigma}^{(n)}_x$ denotes the Pauli-$X$ operator acting on the $n$th qubit. Simultaneously realizing the desired entangling phases $\varphi_{nm}$ for all ion pairs $(n,m)$ requires careful engineering of the dynamical control protocol. In the following, we consider multi-tone pulses whose parameters, such as amplitude, phase, and frequency, can be optimized efficiently and robustly, as demonstrated in Ref. \cite{shapira2023fastdesignscalingmultiqubit}.

We model this setting within the \emph{Lamb–Dicke regime}, where the Hamiltonian of the coupled qubit–phonon system can be written (setting $\hbar=1$) as
\begin{equation}
\begin{aligned}
H &= H_0 + V, \\
H_0 &= \sum_{j=1}^{N} \nu_j \left( a_j^{\dagger} a_j + \frac{1}{2} \right)
+ \sum_{n=1}^{N}\frac{\omega_0}{2} \sigma_z^{(n)}, \\
V &= \sum_{n,j=1}^{N}\eta_j^{(n)} 
\left(\Omega_n(t)\sigma_+^{(n)} + \mathrm{h.c.}\right)
\left( a_j + a_j^\dagger \right).
\label{eq:lsf_H}
\end{aligned}
\end{equation}

The first term in $H_0$ describes the $N$ harmonic phonon modes, labeled by $j$, with creation (annihilation) operators $a_j^\dagger$ ($a_j$) and frequencies $\nu_j$. The second term accounts for the energy splitting $\omega_0$ of the $N$ qubits, where ${\sigma}^{(n)}_z$ is the Pauli-$Z$ operator acting on the $n$th qubit. The interaction term $V$ captures the effect of a coherent drive, which couples the motional and qubit degrees of freedom, and generates spin-dependent force. The time-dependent driving field applied to the $n$th ion is denoted by $\Omega_n(t)$, and $\eta_j^{(n)}$ (the generalized Lamb–Dicke parameter) quantifies the coupling strength between the $j$th motional mode and the drive on the $n$th qubit.

For a drive spectrum chosen symmetrically around $\omega_0$, the time-dependent Schrödinger equation is exactly solvable in the Lamb–Dicke regime. The resulting time-evolution operator can be written as
\begin{align}
U(t) &= \prod_{j=1}^{N} U_{j}(t), \\
U_{j}(t) &= D_{j}\!\left(\sum_{n=1}^{N} \alpha_j^{(n)}(t) {\sigma}^{(n)}_x\right)   
\exp\!\left[i\sum_{n<m} \varphi_{nm}^{(j)}(t){\sigma}^{(n)}_x {\sigma}^{(m)}_x\right],
\label{Eq:unitary_lsf}
\end{align}
where $D_j(\alpha)=\exp\!\left( \alpha a_j^\dagger - \alpha^* a_j \right)$ is the displacement operator of the $j$th mode. The quantity $\alpha_j^{(n)}(t)$ represents the contribution of the $n$th spin to the displacement of the $j$th motional mode and is commonly referred to as the phase-space trajectory, while $\varphi_{nm}^{(j)}(t)$ denotes the contribution of the $j$th mode to the entangling phase between qubits $n$ and $m$, at time $t$. Explicit expressions for $\alpha_{j}^{(n)}(t)$ and $\varphi_{nm}^{(j)}(t)$ can be derived using the Magnus expansion~\cite{shapira2023fastdesignscalingmultiqubit} (see also Appendix~\ref{app:MQ_Magnus}).

Achieving a pure spin–spin interaction at the gate time $t=\tau$ requires spin–phonon disentanglement, i.e., $\alpha_j^{(n)}(\tau)=0$ for all $j,n$, together with the realization of the target phases $\varphi_{nm}$. This generally leads to a set of $N^2$ nonlinear equations~\cite{shapira2020theory}. Nevertheless, optimal MQ gates can be solved for ion crystals containing more than 100 ions~\cite{shapira2023fastdesignscalingmultiqubit}. Using this approach, one can construct MQ gates with simultaneous and programmable all-to-all connectivity and minimal time and laser power.

\section{Framework for noise analysis}
\label{section:noise_sources}

Coherent error sources typically remain within the framework of Eq.~\eqref{Eq:unitary_lsf} and can be quantified, for example, using a filter-function formalism~\cite{kang2023designing}. Incoherent errors are commonly modeled within the Lindblad master-equation formalism under the assumption of Markovian noise~\cite{Srensen2000,korsch2019lindbladdynamicsdampedforced,haddadfarshi2016high,PhysRevA.111.022613,sutherland2022one,ben2020direct,kang2026nongaussianphasetransitioncascade}. 

The Lindblad evolution of a density matrix $\rho$ is given by,
\begin{equation}
\label{Eq:lind}
\begin{aligned}
\dot{\rho}(t)
&= -\frac{i}{\hbar}[H,\rho(t)] + \mathcal{L}(\rho), \\[4pt]
\mathcal{L}(\rho)
&= \sum_n \Gamma_n\!\left(
C_n \rho C_n^\dagger
- \frac{1}{2}\left\{ C_n^\dagger C_n,\rho\right\}
\right),
\end{aligned}
\end{equation}
where $C_n$ depends on the specific noise source and $\Gamma_n$ denotes the corresponding decoherence rate. In this work, we focus on spin dephasing and depolarization, for which $C_n\in\{\sigma_x,\sigma_y,\sigma_z\}$~\cite{ben2020direct} . In addition, we examine motional heating and motional dephasing, described by $C_n\in\{a,a^{\dagger}\}$~\cite{haddadfarshi2016high,Srensen2000} and $C_n=a^{\dagger}a$~\cite{sutherland2022one,kang2026nongaussianphasetransitioncascade}, respectively.
We then characterize the effective noise channel acting on the reduced density matrix of the qubits after tracing out the phonon degrees of freedom.

To simplify the analysis, we consider the evolution 
with respect to the ideal gate $U_{\mathrm{id}}$, by setting
\begin{equation}
\tilde\rho(t) = U_{\mathrm{id}}^{\dagger}(t)\,{\rho}(t)\,U_{\mathrm{id}}(t),
 \label{eq:rho_rotate_id}
\end{equation}
i.e., the noise is factored as acting prior to the ideal evolution,
with $\tilde\rho(t)$ capturing the noise dynamics.  
In this frame, \cref{Eq:lind} becomes
\begin{equation}
\label{eq:lind_rot}
\dot{\tilde{\rho}}(t) = \tilde{\mathcal{L}}\bigl( \tilde{\rho}(t) \bigr),
\end{equation}
where the rotated Lindbladian $\tilde{\mathcal{L}}$ is defined in terms of a trivial transformed Hamiltonian, $\tilde{H}=0$, and  transformed jump operators
\begin{equation}
\tilde{C}_n(t) = U_{\mathrm{id}}^\dagger(t)\, C_n \, U_{\mathrm{id}}(t).
\label{eq:c_transform}
\end{equation}

To leading order in $\Gamma_n\tau$, with $\tau$ the gate duration, the solution of the Lindblad equation takes the form
\begin{equation}
\label{eq:LeadingOrder}
\tilde{\rho}(t)
=
\tilde{\rho}(0)
+\sum_n
\Gamma_n
\int_0^\tau dt
\left[
\tilde{C}_n\tilde{\rho}(0)\tilde{C}_n^\dagger
-
\frac{1}{2}
\left\{
\tilde{C}_n^\dagger\tilde{C}_n,
\tilde{\rho}(0)
\right\}
\right].
\end{equation}

Intuitively, this expression corresponds to a single-event approximation of a Poisson process.  
For unitary jump operators satisfying $C^\dagger C=1$, as in the case of Pauli noise, the anti-commutator term integrates trivially, and the incoherent dynamics arise solely from the first term in the integrand.  

In the following, we assume that the initial state consists of the qubits prepared in the computational-basis ground state and the motional mode in a thermal state, $\bigl(|0\rangle\langle 0|\bigr)^{\otimes N}\otimes\rho_{\text{ph}}$, and consider an MQ gate in the $X$ basis (see Eq.~\eqref{Eq:unitary_lsf}). Under these assumptions, the structure of noise propagation can be inferred directly from the density-matrix elements after the noise has acted.

Our ultimate goal is to derive an effective Pauli-channel noise model suitable as an input for a Clifford-circuit simulation for QEC threshold estimation.  
Although these noise channels may, in principle, generate coherent effects, we assume that such contributions are negligible \cite{Bravyi2018,PhysRevLett.121.190501,Greenbaum2017} because the noise is treated as white and syndrome measurements are performed periodically. Thus accumulation of long-range temporal correlations is suppressed.

\section{Results}
\label{section:analytical}
\subsection{Photon scattering and spin dephasing}
\subsubsection{Single event approximation}
We assume photon scattering rates are low, relative to the gate time (see App.~\ref{app:ramman}). In this limit, noise is modeled by stochastic Pauli errors acting on each qubit independently, generating spin depolarization and spin dephasing channels, described by the application of ${\sigma}^{(k)}_{x,y,z}$ and ${\sigma}^{(k)}_z$ operators on qubit $k$, respectively. Spin depolarization corresponds to processes that randomizes the qubit state within the Bloch sphere and can originate, for example, from spontaneous photon scattering during Raman-driven gates (see App.~\ref{app:ramman} and \cite{ozeri2007errors}). Dephasing can arise as well from photon scattering, but also from magnetic-field fluctuations \cite{Langer2005} or laser phase/frequency noise, which induces stochastic shifts of the qubit transition frequency. Recoil of the qubit due to the scattering is neglected since  in the Lamb-Dicke regime, in which the gate operates, it is suppressed. Nevertheless, the recoil amounts to incoherent displacement of the ions, which is modeled by mode heating, described below.

For both depolarization and dephasing, the appearance of Pauli operators that anti-commute with the gate's Pauli operators reverses $\sigma_x$ to $-\sigma_x$ and therefore reverse the direction of the spin-dependent forces mid-gate. This will lead to non-closure of phase space trajectory at the gate time, i.e. excess displacement which comes about a single-qubit errors, and to an error in the accumulated entanglement phase, i.e. errors in the $\varphi_{nm}$s. 

For the specific choice of Ising gates described in \cref{eq:Umq}, ${\sigma}^{(k)}_x$-errors commute with the operations and hence trivially factor out and generate a local bit-flip channel. With the above discussion in mind and without loss of generality, for the rest of this section we only consider the jump operator $C={\sigma}^{(k)}_z$, acting on the $k$'th qubit. ${\sigma}^{(k)}_y$ operators act analogously.

In the rotated frame of Eq. \eqref{eq:lind_rot}, the jump operator takes the form 
\begin{equation}
\begin{aligned}
\tilde{C}(t) &= U_{\mathrm{id}}^\dagger(t)\, C \, U_{\mathrm{id}}(t) \\
&= D_j\Bigl( -2 \alpha_j^{(k)}(t) \, {\sigma}^{(k)}_x \Bigr) e^{-2 i \sum_{n\neq k}\varphi_{kn}(t) {{\sigma}^{(k)}_x {\sigma}^{(n)}_x}}\sigma_z^{(k)}.
\label{eq:c_scattering_transform}
\end{aligned}
\end{equation}
The above expression results from the anti-commutation relation between the noise term $\sigma_z^{(k)}$ and the operator ${\sigma}_x^{(k)}$ that appears in the ideal gate evolution, $U_{\mathrm{id}}$. 

The rotated jump operator is then used in the leading order expansion of \cref{eq:LeadingOrder}, with scattering rate denoted by $\Gamma^s$ and gate time $\tau$. To isolate the qubit error from phonon error, we consider the reduced density matrix of the qubit subspace by tracing out the phonons. 

We note that only the first factor in $\tilde{C}$, corresponding to the spin-dependent phonon displacement, entangles the spin and phonon degrees of freedom. For this reason, the trace over phonons only receives the following contribution:
\begin{equation}
\begin{aligned}
\mathrm{Tr}_{\rm ph} & \left[ D_j\left( 2 \alpha_j^{(k)}(t) {\sigma}_x^{(k)} \right)\rho_0(0)
D_j\left( -2 \alpha_j^{(k)}(t) {\sigma}_x^{(k)} \right) \right] \\
&= (1-\gamma_k)\, I \, \rho_0(0) \, I 
+ \gamma_k {\sigma}_x^{(k)} \rho_0(0) {\sigma}_x^{(k)}.
\label{eq:phonon_trace}
\end{aligned}
\end{equation}
The above expression describes a bit-flip channel on the $k$'th qubit with probability density per unit time $\gamma_k(t)$, which reads,
\begin{equation}
\label{eq:dephasing_k}
\gamma_k(t) = \frac{1 - \exp\Big[-2\sum_j (2 \bar{n}_j + 1)|\alpha_j^{(k)}(t)|^2\Big]}{2},
\end{equation}
with $\bar{n}_j$ the average thermal occupation of the $j$th phonon.

We thus find that,  to leading order in $\Gamma^s\tau$, the effect of scattering noise is as follows: (i) A bit-flip channel ${\sigma}_x^{(k)}$ localized on ion $k$, (ii) two-qubit over rotations $R_{xx}\left(2 \varphi_{nk}\right)$ for all $n\neq k$ and (iii) a phase-flip $\sigma_z^{(k)}$.

The probability $p_n^{(k)}$ that a specific qubit $n$ was flipped is obtained by tracing out over all other qubits, yielding (see App.~\ref{app:bitflip_analyitcal})
\begin{equation}
\label{Eq:bitflip}
p_n^{(k)}=
\begin{cases}
\Gamma_\text{s}\int_0^{\tau} \sin^2 \big( 2\varphi_{nk}(t) \big) \, dt & n \neq k \\  
\Gamma_\text{s}\int_0^\tau   \big(\frac{1}{2}-\frac{(1-2\gamma_k)}{2}
\prod_{n'\neq k}\cos\!\big(4\varphi_{n'k}(t)\big)\big) dt & n=k .
\end{cases} 
\end{equation}

Importantly, \(\varphi_{nk}(t)\) may be finite at intermediate times $t\in\left[0,\tau\right]$, even in cases where qubits \(n\) and \(k\) are disconnected at gate time (\(\varphi_{nk}(\tau) = 0\)). Nonetheless, minimizing \cref{Eq:bitflip} suppresses error propagation to disconnected qubits.  

Next, we generalize the above relation to also account for MQ correlated errors. Together with the phase-flip term $\sigma_z^{(k)}$, the diagonal elements of the density matrix obey the following noise channel:
\begin{equation}
    \begin{aligned}
        \tilde{\rho}(\tau)=\left( 1-\Gamma_\text{s} \tau \right) I \tilde{\rho}(0)I+\sigma_z^{(k)}\left(\sum_{\boldsymbol{n}} p_{\boldsymbol{n}}^{(k)} P_{\boldsymbol{n}} \tilde{\rho}(0) P_{\boldsymbol{n}} \right) \sigma_z^{(k)}.
    \end{aligned}
    \label{eq:scattering_density_matrix}
\end{equation}

In the above equation, $\boldsymbol{n}$ is a binary vector indicating the presence of either a nontrivial Pauli-$X$ (bit-flip) operation or the identity operator, according to
\[
P_{\boldsymbol{n}}=\bigotimes_i (\sigma_{x_i})^{n_i}.
\]

For each qubit $n\neq k$, a flipped (or non-flipped) qubit contributes a probability factor of $\sin^2(2\varphi_{nk})$ (or $\cos^2(2\varphi_{nk})$) to $p_{\boldsymbol{n}}^{(k)}$. The contribution of the faulty qubit $k$, depends both on the phonon-induced bit-flip probability (i.e., $\gamma_k$) and on the total number of two-qubit flip events induced by over-rotations, since each such event involves the faulty qubit $k$. A more explicit example for the case of a five-qubit MS gate can be found in App.~\ref{app:analytical_MS}.

In general, the multi bit-flip probability given by:
\begin{equation}
\begin{aligned}
p_{\mathbf{n}}^{(k)} &= \Gamma_\text{s} \int_0^{\tau} \!\! dt \!\!\!\!\!\! 
\prod_{s \in \text{supp}\left(P_\mathbf{n}\right)} \!\!\!\!\!\!\sin^2\left( 2 \varphi_{s k}(t)\right) \!\!\!\!\!\!\!\!\!\!\!\!\!\!\!\!\prod_{s' \in \{1..N \} \backslash \text{supp}\left(P_\mathbf{n}\right)} \!\!\!\!\!\!\!\!\!\!\!\!  \cos^2\left( 2 \varphi_{s' k}(t) \right) \\ 
& \times \begin{cases}
    1-\gamma_k(t) & \sum_{i}n_i\  (\text{mod} \ 2) = 0 \\
    \gamma_k(t) & \sum_{i}n_i\  (\text{mod} \ 2) = 1
\end{cases},
\end{aligned}
\label{eq:p_nvec}
\end{equation}
where $\mathrm{supp}(P_\mathbf{n})$ denotes the set of ion indices with $\sigma_x$ support, and its complement, $\{1,\dots,N\} \setminus \mathrm{supp}(P_\mathbf{n})$, denotes the set of indices with trivial support. As implied by the above expression, we note that the bit-flip probability of qubit $k$ depends on the parity of $\mathbf{n}$, since each flip of qubit $s$ is correlated with a flip of qubit $k$. Accordingly, one should choose either $1 - \gamma_k$ or $\gamma_k$ in $p_{\boldsymbol{n}}^{(k)}$. Consistently, $\sum_{\mathbf{n}} p_{\mathbf{n}}^{(k)} = \Gamma^{s}\tau$, such that these processes complement the null event, with probability $1-\Gamma_{\text{s}}\tau$, forming a well-defined quantum channel. As mentioned above, we focus on the diagonal elements only, neglecting coherent (off-diagonal) errors.

The analytical formula for the MQ density matrix is one of our key results, as it provides a simple and tractable estimate for the error in leading order. For instance, it agrees to high precision with numerically exact simulations of scattering errors in the MS gate on 5 qubits, see Ref. \cite{Schwerdt2022} and App.~\ref{app:analytical_MS} for more details.

\subsubsection{MQ gate errors in surface code syndrome extraction}
\label{sec:scatter_surface_code}
\begin{figure*}
    \centering
    \includegraphics[width=1.5\columnwidth]{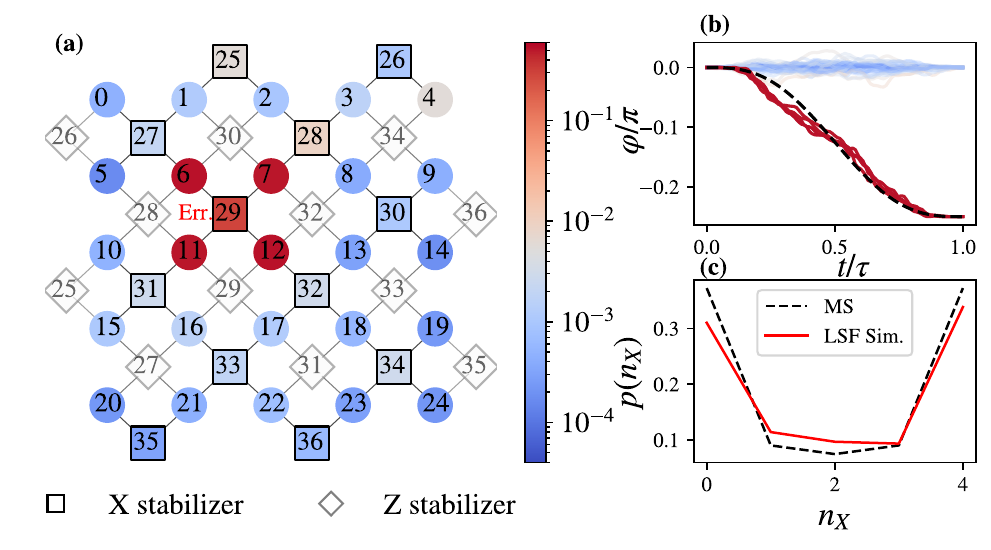}
    \caption{ The effect of Raman scattering on $X$ stabilizers measurements of the distance 5 rotated surface code, implemented with a single MQ gate. 
    (a) bit-flip probability (color, log scale) for each qubit, conditioned on a $\sigma_z$ error on a specific qubit (here the ancilla qubit $q_{err}=29$). Error rates associated with data qubits not directly coupled to the faulty qubit are markedly small, consistent with the connectivity structure imposed by the stabilizer. (b) The entanglement phase trajectories $\varphi_{n,q_{err}}\left(t\right)$ as a function of time ($dt=0.01\tau,\tau \approx 321\text{ }\mu\text{s}$). Colors correspond to overall bit-flip probabilities in panel (a) for each $n\neq q_{err}$. The entanglement phase evolution is compared to the MS phase evolution in Eq.~\eqref{eq:ms_phi} (dashed black), displaying a qualitative fit. (c) The correlated error probabilities (red), where $n_X=0,...,4$ (horizontal) is the number of flipped data qubits, in the $X$ stabilizer measured by the faulty ancilla qubit $q_{err}$. The MS results are for 5-qubit gate where the faulty qubit was traced out.
    }
\label{fig:scattering}
\end{figure*}

To exemplify the above analysis, we study the $d = 5$ rotated surface code shown in Fig.~\ref{fig:scattering}\textcolor{blue}{a}. The code parameters are $[n=25,k=1,d=5]$, requiring $24$ stabilizers. All-to-all connectivity allows recycling of ancilla qubits for measuring $Z$ and $X$ stabilizers, such that only $12$ ancilla qubits are required (see for example~\cite{reichardt2024demonstrationquantumcomputationerror}). For that reason the same ancilla qubit indices are used for $Z$ (diamond) and $X$ (square) stabilizers.

Within our MQ formulation, we consider a gate that performs simultaneous parity checks of all $X$ stabilizers (a similar process can be done for $Z$ stabilizers) .  
This can be implemented by choosing the entangling phases from \cref{Eq:unitary_lsf} as $\varphi_{nm} = \pi/4$ if $\{ n,m\}$ is a pair of ancilla and data qubit belonging to the same stabilizer and otherwise, $\varphi_{nm} = 0$. In Fig.~\ref{fig:scattering}\textcolor{blue}{a} , the lines connecting data qubits (circles) to  $X$  ancilla qubits (squares), represent finite $\varphi_{nm}$. This construction is equivalent to the standard decomposition into a sequence of two-qubit gates~\cite{Schwerdt2022}, up to two layers of single-qubit rotations applied before and after the MQ gate.

The analytic expression in \cref{Eq:bitflip,eq:p_nvec} provides a guiding principle for controlling scattering errors. Explicitly, minimizing $p_{n}^{(k)}$ entails keeping disconnected entangling phases ($\varphi_{nm} = 0$) small throughout the entire gate evolution, and not only at gate time. As typically LSF generates multiple candidate solutions which incur similar drive power and gate fidelity, this criterion above enables to select a sub-set of the solutions in which $p_{n}^{(k)}$ is minimized. We generated an MQ parity check gate informed by the above condition, using LSF optimization \cite{shapira2023fastdesignscalingmultiqubit}.  The results of this analysis are summarized in Fig.~\ref{fig:scattering}, where we focus on the more challenging case of a faulty, highly-connected ancilla qubit. The simpler case of a faulty data qubit is discussed in App.~\ref{app:gate_param}.

In Fig.~\ref{fig:scattering}\textcolor{blue}{a}, we provide a heatmap depicting individual bit-flip probabilities conditioned on a $\sigma_z$ error event occurring on the ancilla qubit $k=29$ (\cref{Eq:bitflip}) during gate evolution. We assume $\bar{n}_j=0$ for all $j$, namely the phonon ground state. 
Our key finding is that only qubits connected to the faulty bit, as dictated by non-vanishing $\varphi_{nm}$, admit sizable bit-flip probabilities. In particular, the bit-flip probability for qubits that are disconnected from the faulty qubit is roughly two orders of magnitude smaller. Therefore, the spatial error propagation is \emph{localized} to the support of $\varphi_{nm}$.

The locality of the error follows directly from the $\varphi_{nk}(t)$ trajectories, shown in Fig.~\ref{fig:scattering}\textcolor{blue}{b}. For qubits not directly connected to the faulty ancilla, $\varphi_{nk}(t)$ remain small throughout the \textit{entire} evolution interval $t \in [0,\tau]$, and not merely at  $t=\tau$, where the desired zero connectivity is realized.
As noted above, and evident from \cref{Eq:bitflip}, this requirement is essential to prevent error propagation.
Indeed, if the MQ gate design results in $\varphi_{nm}(t)$ that displays large deviation from zero during the gate evolution (even if 
$\varphi_{nm}(t=\tau)=0$) it leads to undesired error propagation.  An illustrative case for such error prone gate design is analyzed in Appendix~\ref{app:gate_param}.

Going beyond single-bit-flip errors, we now turn to analyzing MQ-correlated errors. To that end, we compute the probability of correlated errors on the four data qubits connected to the faulty ancilla qubit $k$ via $\varphi_{nk}$ after tracing out the ancilla itself, see red curve in Fig.~\ref{fig:scattering}\textcolor{blue}{c}. Crucially, almost $\sim 95\%$ of the entire probability mass is concentrated on the $2^5$ (to be compared with the full $2^{37}$) dimensional Hilbert space defined on qubits belonging to the support of non-vanishing $\varphi_{nk}$. This further substantiates that error propagation through the MQ gate remains localized. 

An additional key observation is that the most dominant error channel is a correlated four-qubit flip.
Interestingly, from the point of view of error correction, such an error is trivial up to a stabilizer. 
By contrast, parity measurements based on a discrete sequence of two-qubit CNOT gates suffer from hook errors, as each Pauli string error channel appears with an equal probability.
See Appendix~\ref{app:analytical_MS} for a more detailed comparison of the hook-error effect between two-qubit gates and MQ gates.

It is illuminating to compare the above results with the standard MS gate. Motivated by the locality of scattering errors in MQ gates, we consider an MS gate driving the center-of-mass mode and acting \emph{exclusively} on a five-qubit subset comprising the four data qubits and the ancilla qubit. In this simplified scenario, one can analytically compute the evolution of entangling phases, given by
\cite{Srensen1999,Srensen2000}:

\begin{equation}
    \varphi_{nk}(t)=\frac{\xi t - \sin(\xi t)}{8}, \qquad \xi = 2\pi/\tau.
\label{eq:ms_phi}
\end{equation}

Substituting the above expression allows for an exact evaluation of the correlated error probabilities appearing in \cref{eq:p_nvec}, see App.~\ref{app:analytical_MS}. More broadly, we found qualitative agreement between the typical entangling-phase evolution of MQ gates, generated via the LSF solution, and that of the canonical MS gate, see Fig.~\ref{fig:scattering}\textcolor{blue}{c}. This allows us to approximate MQ gate error rates using the simpler analytical expression for the MS gate.

Utilizing the above observations allows for a faithful noise modeling of Raman scattering and spin dephasing to leading order. As we detail in \cref{section:QEC}, the resulting error rates serve as an input for QEC analysis in the context of the threshold transition of the surface code. 

\subsection{Motional heating and dephasing in MQ gates}
\label{sec:x_dephasing}
Ion-crystals are formed by a combination of controllable electromagnetic fields and the Coulomb electric field. Expanding about the equilibrium ion configurations in the presence of trapping potential yields the global vibrational modes of the ion-crystal. 
These phonons mutually interact with all qubits involved in the MQ gate and can therefore potentially introduce correlated errors. It is thus crucial to understand their effect, especially in the context of QEC, where such correlations negatively affect the performance.
We are mainly interested in heating and motional dephasing, as these processes have a leading effect in standard two-qubit gate schemes \cite{Ballance2016}. In our analysis, we assume an initial thermal phonon state for each gate, even when considering gate concatenation. 

Below, we show that if the MQ gate operates in the $X$ basis both effects produce solely X-dephasing channels on the qubit subsystem after tracing out the motional degrees of freedom. Consequently, for a general initial density matrix, since the ideal gate and the error channels are diagonal in the $X$ basis, one can parametrize the rotated density matrix at gate time as:
\begin{equation}
\label{eq:x_channel}
\tilde\rho(\tau) = \sum_{ij} \eta_{ij}(\tau)\, P_i\, \tilde\rho(0)\, P_j.
\end{equation}
\(P_i\)s are Pauli strings containing only $I$ and $\sigma_x$ operators. $\eta_{ij}(\tau)$ is a completely positive trace-preserving (CPTP) process matrix, to be determined below, for specific cases. We note that the diagonal entries of $\eta(\tau)$ represent bit-flip probabilities.
To read out the elements of $\eta_{ij}(\tau)$, it is useful to consider the specific initial density matrix $\bigl(|0\rangle\langle0|\bigr)^{\otimes N}\otimes \rho_{\text{th}}$. In this setting, we obtain (see App.~\ref{app:eta_ij}):
\begin{equation}
\eta(\tau) = H^{\otimes N} \tilde\rho_{x,x'}(\tau)\, H^{\otimes N},
\label{eq:had}
\end{equation}
with $H$ being the Hadamard transform and $\tilde\rho_{x,x'}(\tau)$ are the x-basis elements of the density matrix at gate time in the rotated frame.

\subsubsection{Motional heating}

In the following, we generalize the treatment of thermal effects in Ref.~\cite{Srensen2000} to account for multiple phonon modes. Intuitively, mode heating generates an excess displacement of the phonon modes, disturbing the closure of the phase-space trajectory at the gate time. In leading order this manifests as as residual qubit-phonon coupling, which is a single-qubit error, and in second order an over-rotation of the entanglement operation. We analyze the Lindblad evolution of \cref{Eq:lind}, taking the jump operators for $j$'th mode:
\begin{equation}
C_{1,j} = \sqrt{\Gamma_{\text{h},j}(1+\bar{n}_{\text{th},j})}\, a_j, \qquad 
C_{2,j} = \sqrt{\Gamma_{\text{h},j} \bar{n}_{\text{th},j}}\, a_j^{\dagger},
\label{eq:jump_heating}
\end{equation}
with $\bar{n}_{\text{th},j}$, $\Gamma_{\text{h},j}$ are the steady state's thermal occupation and the characteristic relaxation rate of the $j$'th mode, respectively. Following the procedure in \crefrange{eq:rho_rotate_id}{eq:c_transform}, the rotated jump operators of the $j$'th mode take the form:

\begin{equation}
\label{eq:c_heating}
\begin{split}
\tilde{C}_{1,j} &= \sqrt{\Gamma_{\text{h},j}(1+\bar{n}_{\text{th},j})} \left(a_j + \sum_{n} \alpha_j^{(n)}(t) {\sigma}_x^{(n)}\right),
\end{split}
\end{equation}

\begin{equation}
\begin{split}
\tilde{C}_{2,j} &= \sqrt{\Gamma_{\text{h},j} \bar{n}_{\text{th},j}} \left( a_j^\dagger + \sum_{n} \alpha_j^{*(n)}(t) {\sigma}_x^{(n)} \right).
\end{split}
\end{equation}

 The Lindblad equation can be solved analytically (see Ref.~\cite{Srensen1999}). Upon tracing out the phonons, the rotated density matrix $\tilde{\rho}(t)$ projected to the $X$ basis, obeys the differential equation:

\begin{equation}
\begin{split}
\label{eq:heating_rho_ode}
\dot{\tilde{\rho}}_{\boldsymbol{x}',\boldsymbol{x}} 
=&-\sum_{j}\frac{\Gamma_{\text{h},j}(2\bar{n}_{\text{th},j}+1)}{2}\left|\sum_{n} \alpha_j^{(n)}(t)\, (x_n-x'_n)\right|^2 
\!\!\!
\tilde{\rho}_{\boldsymbol{x}',\boldsymbol{x}} .
\end{split}
\end{equation}
Here, $\boldsymbol{x}$ is a bit string whose elements $x_n=+1$ ($x_n=-1$) correspond to the $x$-basis configuration in which the $n$th qubit is in the $\ket{+}$ ($\ket{-}$) state.

For simplicity, we focus on the MS center-of-mass mode case in the following analysis. This choice is also physically justified, since typically the center-of-mass (COM) mode exhibits the dominant heating rate, as it couples most strongly to electric-field noise~\cite{brownnutt2015ion}. This analysis holds in the general multi-mode case by introducing mode-specific heating rate, $\Gamma_{\text{h},j}$, and weighting each ion according to its mode-dependent displacement, $\alpha_j^{(n)}$ (see App.~\ref{app: heating_LSF}).

Integrating \cref{eq:heating_rho_ode}  for the MS gate and the initial state $\tilde{\rho}(0)=\left(|0\rangle\langle0|\right)^{\otimes N}\otimes \rho_{\text{th}}$  yields (Ref.~\cite{Srensen2000}):
\begin{equation}
\label{eq:rho_ms}
\tilde\rho_{\boldsymbol{x}',\boldsymbol{x}} (\tau)
= \frac{1}{2^N} \exp\left[-\frac{1}{4} (2\bar{n}+ 1)\Gamma_{\text{h}}\tau
\left|\sum_n (x_n - x'_n)\right|^2 \right]
\end{equation}
where we have omitted the $j$ indices as we are only treating the COM mode.

We note that a heating event is accompanied by a global operation on all spins, as seen in Eq. \eqref{eq:c_heating}, with off-diagonal contributions to $\eta$. As is standard, we will focus solely on the diagonal part of $\eta$ to analyze QEC performance. In particular, for diagonal elements, each heating event occurs with probability $\Gamma_{\text{h}}\tau$ and produces a displacement term $\sim \sum_n \sigma_x^{(n)}$, seen in Eq. \eqref{eq:c_heating}, corresponding to a \textit{single bit-flip} event. 
To leading order this yields 
\begin{equation}
    \tilde{\rho}_\text{1q}=\sum_{j,n}\Gamma_{\text{h},j}\int_0^\tau dt  \left|\alpha_j^{(n)}(t)\right|^2(2\bar{n}_j+1) \sigma_x^{(n)}\tilde{\rho}(0)\sigma_x^{(n)}.
\label{eq:dephasing_1q}
\end{equation}

MQ bit-flip events are associated with higher order corrections to the error, beyond leading order in $\Gamma_{\text{h}}\tau$. We show in App.~\ref{app: heating_LSF} that the diagonal elements of $\eta(\tau)$ decay exponentially with the Hamming weight of states they represent, also for a general MQ gate.

\begin{figure*}
    \centering
\includegraphics[width=1.7\columnwidth]{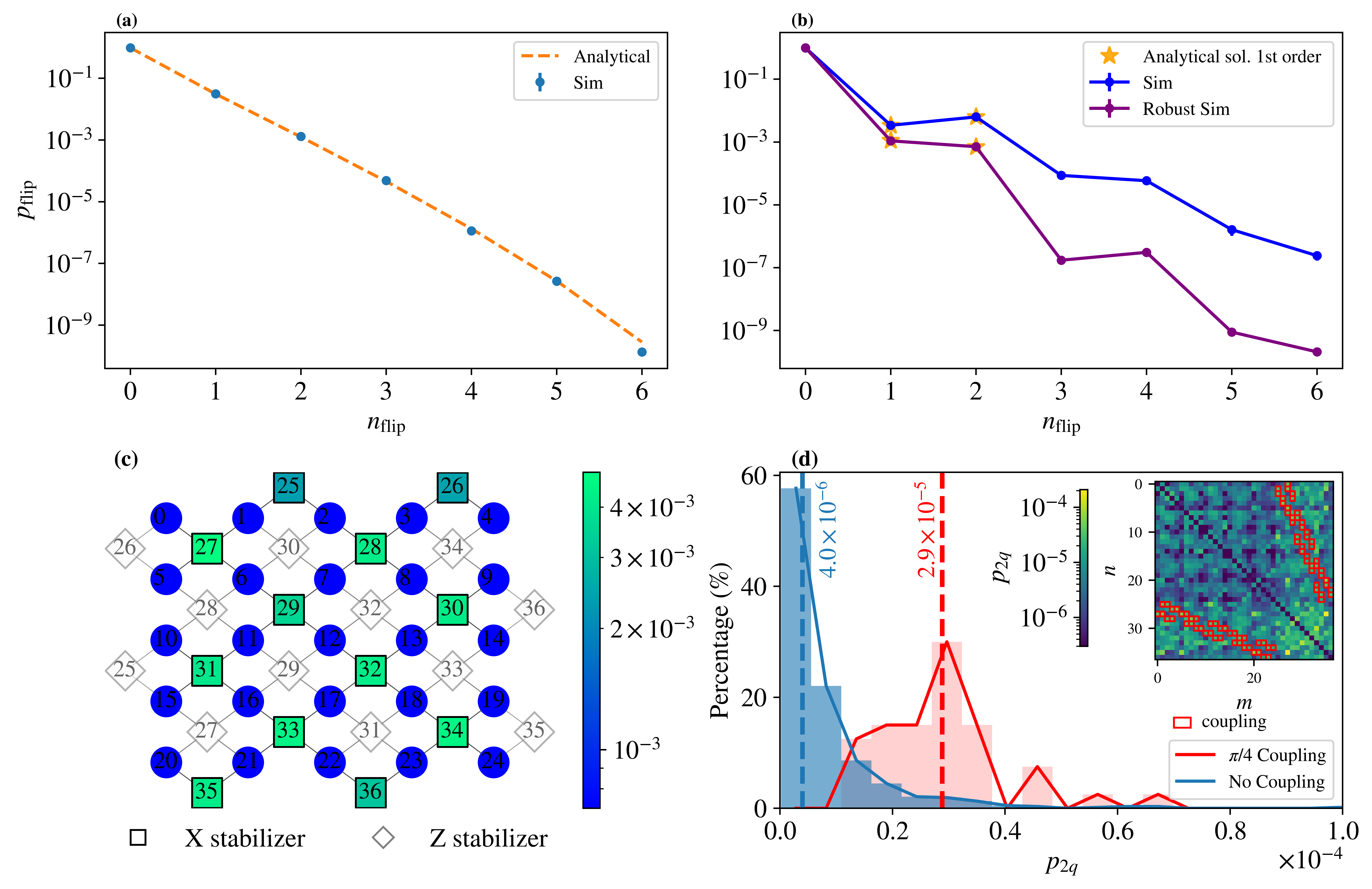}
    \caption{Heating and motional dephasing noise analysis. All values of $\Gamma_{h},\Gamma_{s}$ mentioned below are in-line with realistic hardware performances.
(a) Bit-flip probabilities induced by heating during a MS gate ($N=6$), for $\Gamma_{h} n_{\mathrm{th}} = 50$Hz and $\tau = 450 \,\mu\mathrm{s}$. We observe excellent agreement between the analytical solution and exact numerical simulations. Heating is dominated by single-qubit error with probability $\sim \Gamma_{h} n_{\mathrm{th}} \tau$. 
(b) Bit-flip probabilities induced by motional dephasing for an MS gate ($N=6$), with $\Gamma_{d} = 10$Hz and $\tau = 450\,\mu\mathrm{s}$. Numerical results agree with the leading-order analytical solution (orange stars). Motional dephasing leads to both single- and two-qubit errors. Robust control techniques significantly reduce the error rates. 
(c) Bit-flip probabilities per qubit due to heating for the gate in \cref{fig:scattering} with $\tau \approx  321\,\mu\mathrm{s}$ and $\Gamma_{h} = 100$Hz. Data qubits exhibit lower error rates than the ancilla qubits (in the bulk), as they are coupled to fewer qubits and experience reduced effective drive strength. 
(d) Motional dephasing for the gate in \cref{fig:scattering} with $\Gamma_{d} = 100$Hz. The histograms show the distribution of two-qubit error of coupled and uncoupled entries in the connectivity matrix (each histogram was normalized by the total number of coupled\textbackslash uncoupled entries). The dashed lines indicate the median values. Connected entries ($\varphi_{nm} = \pi/4$) exhibit a median error rate approximately an order of magnitude larger than uncoupled ones. The inset shows the error rates for any $n,m$ in the connectivity matrix, where red squares mark connected entries. 
    }
    \label{fig:n6}
\end{figure*}

For concreteness, we demonstrate the above findings (Eq.~\eqref{eq:had} and Eq.~\eqref{eq:rho_ms}) for the MS gate acting on a $N=6$ ion register with error rate  $\Gamma_{\text{h}} \bar{n}_\text{f}\tau= 2.25\times 10^{-2}$. In Fig.~\ref{fig:n6}\textcolor{blue}{a}, we present the bit-flip error probability (dashed orange) as a function of the number of flipped bits at gate time. We compare our findings to an exact solution of the Lindblad equation (\cref{Eq:lind}) using quantum Monte-Carlo simulation \cite{johansson2012qutip} (blue circles).
We used $5\times10^{4}$ iterations (statistical error bars are shown, though they are small) and a large phonon cutoff, $n_{\mathrm{max}}=40$, in order to obtain a good fit even for rare events.
As expected, we observe a clear exponential suppression in the bit-flip probability. 
Note that even though the noise channel acts globally on all spins, it is weighted with the prefactor, $\left(\alpha_j^{(n)}\right)^2 \sim 1/N$, due to the uniform participation of ions in the COM mode. Consequently, overall there is no scaling with $N$, such that for the sake of QEC protocols, heating can be thought of as a classical single spin-flip.

\subsubsection{Motional dephasing}

Fluctuations in the trapping potential cause correlated drifts in the normal modes frequencies. The gate error depends sensitively on the typical drift timescale, compared with gate duration. Slow drifts produce coherent errors that can be typically mitigated by optimal control techniques \cite{shapira2018robust,jia2023angle}. By contrast, fast drifts, coined motional dephasing, pose a significant challenge and are analyzed below using the Lindblad formalism \cite{sutherland2022one,kang2026nongaussianphasetransitioncascade}. 

Intuitively, a shift in the normal mode frequency will be accumulated throughout the gate evolution and cause residual qubit-motion coupling as well a non-negligible accumulation of excess entanglement, i.e. errors in the $\varphi_{nm}$s. Both effects are relevant in leading order.

The Lindblad jump operator corresponding to motional dephasing is,
\begin{equation}
    C_j = a_j^\dagger a_j,
\end{equation}

with a decoherence rate $\Gamma_{\text{d},j}$~\cite{Ballance2016}. Following the same procedure outlined above, we rotate the jump operator with respect to the ideal unitary evolution:

\begin{equation}
 \tilde{C_j}= \left(a_j^\dagger + \sum_{n} \alpha_j^{*(n)}(t)\sigma_x^{(n)}
\right)\left(a_j + \sum_{n} \alpha_j^{(n)}(t)\sigma_x^{(n)}\right).
\label{eq:ctilde_dephasing}
\end{equation}

We substitute the rotated jump operators in \cref{eq:LeadingOrder}, to obtain the leading order contribution in $\Gamma_{\text{d},j}\tau$. Per our procedure, we trace out the phonon degrees of freedom and retain only diagonal terms. This results in both single-qubit and two-qubit error terms.
The former takes the same form as in \cref{eq:dephasing_1q}.
Two-qubit errors assume the form:

\begin{equation}
\begin{split}
\tilde{\rho}_\text{2q}\left(\tau\right)&=
    \sum_{j,n,m}\Gamma_{\text{d},j}\int_0^\tau dt \left|\alpha_j^{(n)}(t)\right|^2 \left|\alpha_j^{(m)}(t)\right|^2\\&\cdot\sigma_x^{(n)}\sigma_x^{(m)}\tilde{\rho}(0)\sigma_x^{(n)}\sigma_x^{(m)}.
\end{split}
\label{eq:dephasing_2q}
\end{equation}

This contribution arises from the product of two spin operators in \cref{eq:ctilde_dephasing}. A correlated flip of qubits $n$ and $m$ is proportional to the product $\left|\alpha_j^{(n)}\right|^2\left|\alpha_j^{(m)}\right|^2$. We note that two-qubit errors may occur between any qubit pair that experiences a non-vanishing coupling to the collective phonon motion during gate time. By contrast with the single-qubit error channel, the two-qubit case is independent of the initial phonon thermal occupation. 

To gain intuition, we apply the above analysis to the N-qubit MS gate coupled to the COM mode \cite{Srensen2000}. The resulting single- and two-qubit flip
probabilities,
 assuming $\bar{n}_j=0$ for all $j$ (phonon ground state), are:
\begin{equation}
\begin{split}
    p_1^{\text{MS}}&=N\Gamma_{\text{d}}\int_0^\tau dt \left|\frac{1-e^{2\pi it/\tau}}{4}\right|^2=\frac{N\Gamma_{\text{d}}\tau}{8} \\
    p_2^{\text{MS}}&=2\binom{N}{2}\Gamma_{\text{d}}\int_0^\tau dt \left|\frac{1-e^{2\pi it/\tau}}{4}\right|^4=\binom{N}{2}\frac{3\Gamma_{\text{d}}\tau}{32}.
\end{split}
\label{eq:theory_ms_dephasing}
\end{equation}

We now turn to corroborate the above analysis by benchmarking our prediction with numerically exact simulations. To that end, we simulate the Lindblad equation of the ideal gate with motional dephasing at the rate $\Gamma_{\text{d}}=10\text{Hz}$, $N=6$ and gate time $\tau=450\text{ }\mu\text{s}$ (using quantum Monte Carlo simulations with the same phonon cutoff and number of iterations as for heating). 
Crucially, as seen in ~\cref{fig:n6}\textcolor{blue}{b}, the leading-order contributions in $\Gamma_{\text{d}}\tau$ dominate the single- and two-qubit flipping probabilities, and are in agreement with the theoretical predictions in \cref{eq:theory_ms_dephasing}. Furthermore, error channels involving more than two qubits are suppressed by higher powers of $\Gamma_{\text{d}}\tau$. 

Natural mitigation strategies to cope with motional dephasing involve cooling to the phonon ground state and maintaining phonon trajectories with small displacements throughout the gate evolution. This motivates constraining the phase space trajectories, $\alpha_j^{(n)}$, to remain close to the origin. To that end, one can constrain the maximal magnitude of the $\alpha_j^{(n)}$s, which is a by-product of using low-power solutions of \cref{Eq:unitary_lsf}, shown in Ref. \cite{shapira2023fastdesignscalingmultiqubit} and utilized here. Another typical choice is to use gate designs that constrain the average phase space trajectory around the origin $\int\alpha_j^{(n)}=0$ \cite{shapira2018robust,shapira2020theory}, which typically acts to reduce its magnitude. Applying this constraint to the MS gate, we obtain the bit-flip probabilities: 
\begin{equation}
\begin{split}
p_1^{\text{Robust}}&=N\Gamma_{\text{d}}\int_0^\tau dt \left|\frac{e^{4\pi it/\tau}-e^{2\pi it/\tau}}{\sqrt{24}}\right|^2=\frac{N\Gamma_{\text{d}}\tau}{12} \\
    p_2^{\text{Robust}}&=2\binom{N}{2}\Gamma_{\text{d}}\int_0^\tau dt \left|\frac{e^{4\pi it/\tau}-e^{2\pi it/\tau}}{\sqrt{24}}\right|^4=\binom{N}{2}\frac{\Gamma_{\text{d}}\tau}{96}.
\end{split}
\label{eq:theory_nuoid_dephasing}
\end{equation}
Notably, the resulting two-qubit error probabilities are an order of magnitude smaller than the original MS gate. Moreover this robustness also decreases the relative occurrence of long-range two-qubit errors as the ratio $p_2/p_1$ for the robust gate is smaller by a factor of 6 compared to the MS gate. 

To generalize the above treatment to MQ gates, we use the parameters of the gate presented in \cref{fig:scattering}. We substitute these parameters into \cref{eq:dephasing_1q}  and present the resulting single-qubit error rates for each qubit in \cref{fig:n6}\textcolor{blue}{c}. The data qubits, as well as most of the boundary ancilla qubits, exhibit lower error rates since they are coupled to only two qubits rather than four, in contrast to bulk ancilla qubits. This arises from their lower effective power.
A similar procedure is applied to evaluate two-qubit errors, shown in \cref{fig:n6}\textcolor{blue}{d}. In the inset, we display the values of $p_{2q}$ for all qubit pairs $(n,m)$, where coupled qubits ($\varphi_{nm}=\pi/4$) are highlighted by red rectangles. Additionally, we plot histograms for the coupled entries  and uncoupled entries ($\varphi_{nm}=0$), each independently normalized to $100\%$. We find that the median value of the coupled entries is roughly an order of magnitude larger. This effect can naturally be attributed to the stronger correlations required between the contributions of coupled qubits $n,m$ to the mode displacements, $\alpha^{(n,m)}_j(t)$, that ultimately generate the desired entangling phase (see App.~\ref{app: heating_LSF}). The above observation suggests that a justified simplification of the error model can be made by assuming finite two-qubit errors only between directly coupled qubits, while neglecting two-qubit errors between uncoupled qubits.

\section{Surface-Code Benchmark}
\label{section:QEC}
Combining the effective MQ-gate noise channels derived above, we now turn to assess the required physical error rates for a stable QEC. For concreteness, we simulate the rotated surface code, following the convention illustrated in Fig.~\ref{fig:scattering}, namely taking $d^2$ data qubits and $(d^2-1)/2$ ancilla qubits, with an odd $d\geq5$.

\begin{figure*}
    \centering
    \includegraphics[width=0.9\textwidth]{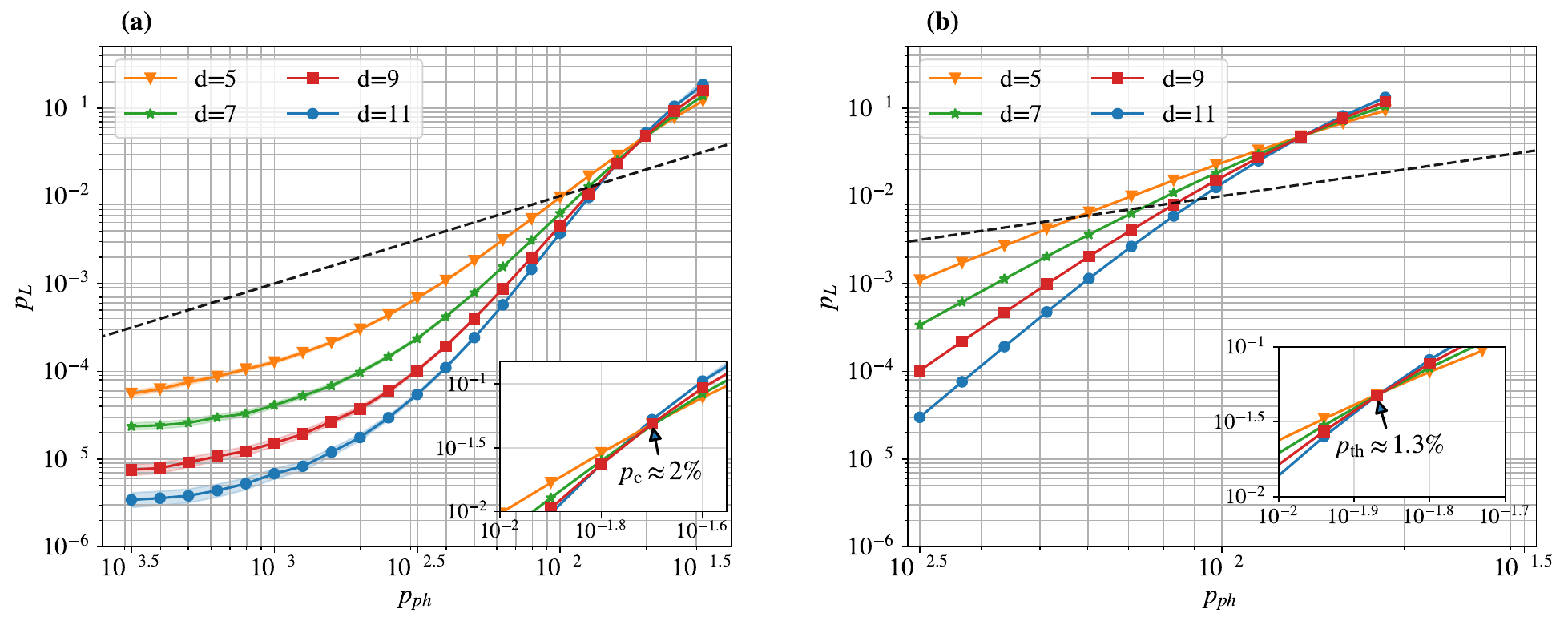}
    \caption{
    Logical error rate as a function of the physical error rate, where all \(X\) (or \(Z\)) stabilizers are measured within a single MQ gate. Two-qubit motional-dephasing errors are introduced in two ways:
In (a), a constant two-qubit error rate of \(p_{2q}=3\times10^{-5}\) is assigned to every pair of qubits, while single-qubit error sources (including idle errors, heating, and scattering) are varied simultaneously with a common physical error rate \(p_{\mathrm{ph}}\). In this case, we observe a crossing at \(p_c \approx 2 \times 10^{-2}\). The inset shows a magnified view of the crossing region.
In (b), two-qubit errors act only between directly coupled qubits. Here, all noise sources, including two-qubit errors, are varied simultaneously with the same \(p_{\mathrm{ph}}\). In this case, a clear threshold is observed at \(p_{\mathrm{th}} \approx 1.3\%\). 
}
    \label{fig:surface_code_qec}
\end{figure*}

Specifically, we consider the following noise sources:
\begin{enumerate}
\item     \textbf{Scattering error:} Modeled by a single-qubit depolarization on all qubits participating in the gate operation, with error propagation probabilities to all qubits coupled to the faulty qubit, according to \cref{eq:p_nvec}.
\item \textbf{Single-qubit errors:} Implemented on all qubits participating in the gate. This error can arise from heating, (\cref{eq:rho_ms}), or mode dephasing (\cref{eq:dephasing_1q}). This error channel acts as a Pauli $X$ ($Z$) when measuring $X$ ($Z$) stabilizers.
\item \textbf{Two-qubit errors:} Implemented between pairs of qubits participating in the gate. These errors arise from motional dephasing (see \cref{eq:dephasing_2q}) and act as $\sigma_{x}^{(i)}\otimes \sigma_{x}^{(j)}$ ($\sigma_{z}^{(i)}\otimes \sigma_{z}^{(j)}$) when measuring $X$ ($Z$) stabilizers.
Below, we consider two models. In the first, we adopt a conservative approach using a uniform two-qubit error rate $p_{2q}$ for any pair $i,j$ of participating qubits,  taken as the median value of the coupled entries ($\approx 3\times10^{-5}$; see the red histogram in \cref{fig:n6}\textcolor{blue}{d}). Alternatively, we consider a model in which a finite $p_{2q}$ is only assigned to directly coupled qubits, while the low error rates of uncoupled pairs are neglected. These two approaches are presented in \cref{fig:surface_code_qec}\textcolor{blue}{a} and \cref{fig:surface_code_qec}\textcolor{blue}{b}, respectively.
\item \textbf{Idle single-qubit depolarization noise:} Determined only by coherence time and independent of the gate dynamics. This quantity serves as a reference rate for physical memory errors. 
\end{enumerate}

We note that, although we do not account for measurement errors, ancilla qubits are still exposed to idle errors and other noise mechanisms described above. We anticipate that measurement noise will only have a quantitative effect on the threshold rate. Our main goal is to demonstrate a physically relevant threshold error rate (or practical crossing point) under the aforementioned noise model, which includes single-qubit errors and multiple sources of correlated errors. Our decoding scheme is the standard minimum-weight perfect-matching decoder \cite{fowler2012surface}. 
Additional details regarding the QEC simulations may be found in App.~\ref{app:qec_details}.
More involved QEC codes and decoders tailored to all-to-all connectivity and MQ gates, may achieve significantly improved performance \cite{self2024protecting,reichardt2024demonstrationquantumcomputationerror,bravyi2024high,nx6p-hjqy,PhysRevLett.133.240602}. Exploring such optimized schemes is left for future work

The results of this analysis are summarized in \cref{fig:surface_code_qec}. In \cref{fig:surface_code_qec}\textcolor{blue}{a}, a fixed two-qubit error probability of $p_{2q}=3 \times 10^{-5}$ was assigned between all pairs of qubits participating in the gate (regardless of whether they are coupled at gate time). The simulation is performed for increasing code distances ($d=5,7,9,11$), and the resulting logical error probability is plotted as a function of the single-qubit error rate.
We observe a clear crossing at $p_{c}\approx 2\times 10^{-2}$, below which logical error rates systematically decrease as a function of code distance.
This constitutes an important result of this work, namely the demonstration of stable error correction within the MQ gate protocol.
We emphasize that $Z$ or $X$ parity check are carried out simultaneously, fully utilizing the advantage of programmable MQ gates in trapped ion systems. 
In \cref{fig:surface_code_qec}\textcolor{blue}{b}, motivated by the results in \cref{fig:n6}\textcolor{blue}{d}, we assume $p_{2q}$ is finite only for directly coupled qubits, i.e, the ancilla and its corresponding data qubits in the relevant stabilizer. In this case, we observe a threshold transition at $p \approx 1.3\%$. 

\begin{figure}[h]
    \centering
    \includegraphics[width=0.49\textwidth]{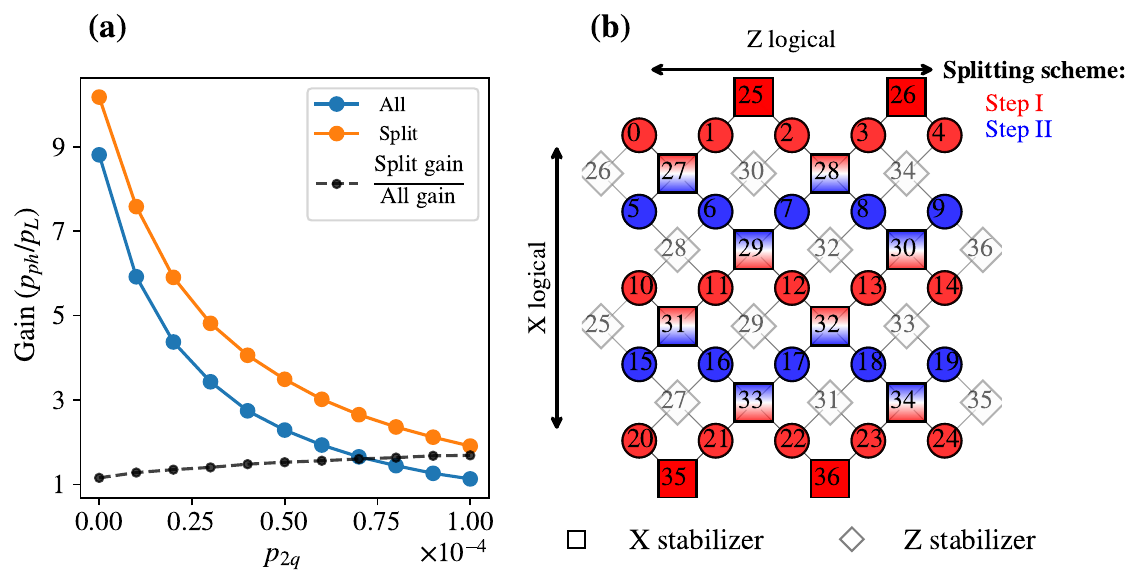}
    \caption{Studying QEC performance as a function of two-qubit error. 
(a) Gain factor ($p_{ph}/p_L$) as a function of $p_{2q}$, with all single-qubit error rates (including idle errors) fixed at $p_{1q}=10^{-3}$ and code distance $d=5$. We compare two measurement schemes: simultaneous measurement of all stabilizers (blue) and a scheme that splits measurements into even and odd rows (orange). For $p_{1q}=10^{-3}$, the splitting scheme yields higher gain. The relative advantage of the splitting protocol, $\frac{\text{Split gain}}{\text{All gain}}$, grows with $p_{2q}$ (dashed black line).
(b) Illustration of the $X$-stabilizer measurement protocol, where measurements are performed in two steps on odd and even rows (perpendicular to the logical $X$, which spans a column) to mitigate the effect of two-qubit errors. The ancilla qubits in the bulk are participating in both steps. The optimal level of parallelism depends on the relative magnitudes of idle and two-qubit error rates.}
    \label{fig:qec_gain}
\end{figure}

As observed, including two-qubit errors between uncoupled qubits at gate time poses a greater challenge and therefore require particular care. In practice, although large two-qubit error rates may degrade error-correction performance with a single MQ gate, a more serial syndrome-extraction scheme can mitigate these effects and improve overall error-correction performance.
To study this effect, we fix all single-qubit error sources to a physically relevant value of $p=10^{-3}$ and measure the error rate gain as a function of $p_{2q}$ for $d=5$ (for any pair of qubits participating in the gate, irrespective of direct coupling).
 In \cref{fig:qec_gain}\textcolor{blue}{a}, we compare the gain factor, $p_{\mathrm{ph}}/p_{L}$, for two syndrome-extraction protocols: (i) measuring all $X$ (or $Z$) stabilizers using a single MQ gate, as in the previous sections, and (ii) splitting the measurement into odd and even rows (columns) for $X$ ($Z$) stabilizers (see illustration of the splitting scheme in \cref{fig:qec_gain}\textcolor{blue}{b}), though still using MQ gates.
 Performing stabilizer measurements sequentially along the direction perpendicular to the corresponding logical operator, rather than executing all MQ interactions simultaneously, can mitigate the effective reduction in code distance induced by two-qubit errors and thereby improve QEC performance~\cite{fowler2012surface}. 
 
 For this choice of parameters, we find a performance gain up to $p_{2q}<10^{-4}$. Notably, the splitting protocol provides a sizable improvement over a single-gate operation. Moreover, the advantage of the splitting protocol becomes more pronounced at larger $p_{2q}$ (dashed black line). The optimal degree of parallelism depends on the interplay between gate error rates and idle error rates.
Our result demonstrates that our trapped-ion motivated noise model supports quantum error correction with realistic, near-term error rates, confirming the practical relevance of MQ gates for QEC. 

\section{Discussion and outlook}
\label{section:discussion}
This work presents a microscopic noise modeling for MQ gate schemes in the context of trapped-ion systems. Integrating out the phonon degrees of freedom offers a practical error model captured via simple Pauli channels. Computing the leading-order contributions in each physical noise channel provides crucial insights into the circuit-level noise model. Notably, via a suitable measurement scheme, we observe a clear QEC gain for the canonical surface code, and an experimentally relevant threshold transition under more relaxed - yet physically justified - assumptions. Below, we summarize and discuss a few key insights:

Scattering errors give rise to MQ error propagation that is restricted to qubits directly coupled to the faulty qubit. Interestingly, for the case of stabilizer measurement, the error probability is concentrated in error channels that are trivial up to a stabilizer. This is a unique feature of the MQ gate, arising from the simultaneousness and continuous phase accumulation. This should be contrasted with the most standard case of two-qubit gates, in which entanglement is built through a discrete sequence of two-qubit gates that are more susceptible to hook errors, potentially degrading QEC performance.

We generalize the effect of phonon heating beyond the conventional two-qubit MS gate to the MQ case. Despite the high connectivity, the resulting error channel remains restricted to local single-qubit errors. On the other hand, we identify motional dephasing as the most challenging error source in the context of MQ gates, as it leads to both single- and arbitrary-range two-qubit errors. Nevertheless, these errors are substantially more likely between directly coupled qubits than between qubits that acquire no entangling phase at gate time.
Such errors are controlled by the magnitude of the phonon displacement, suggesting mitigation schemes, such as the protocol in Ref.~\cite{shapira2018robust}. 

All analyzed noise sources are intuitively understood from their effects on phase-space trajectories. This principle can straightforwardly be generalized to other types of noise, e.g., slow drifts in the drive power, which will dilate the phase space trajectories about the origin; however, at the gate time, they will nevertheless close. Thus, in leading order, this error generates two-qubit errors exclusively among connected qubits.

Looking to the future, we highlight several extensions of our work:

\begin{itemize}

\item \textbf{High rate quantum error correction codes}-High-rate LDPC codes rely on non-planar, non-local connectivity and are therefore naturally suited to trapped-ion architectures that exploit all-to-all connectivity. It would be of great interest to assess the advantages of the MQ gate in the context of high-rate codes with sparse but nonlocal stabilizers, such as in Refs.~\cite{Ye_2025,rmy6-9n89}.

\item \textbf{Optimized decoding}-  It is necessary to consider decoders that explicitly account for correlated noise, such as hypergraph decoders \cite{delfosse2023splitting}, rather than assuming only single-qubit errors as in MWPM.  
 
\item \textbf{Utilizing architecture features}-
It would be interesting to investigate the optimal number of ancilla qubits, which may be reused at the cost of longer syndrome extraction times, as discussed in Ref.~\cite{Ye_2025}. Reducing the number of ancilla qubits without introducing additional swap operations is a key advantage of all-to-all connectivity. Furthermore, we have not considered the ordering of ions within the chain, which can also have a significant impact, as demonstrated in Ref.~\cite{Trout2018}.
\end{itemize}

\begin{acknowledgments}
We would like to thank Erez Berg, David Schwerdt, Lee Peleg and Gilad Kishony for enlightening and valuable discussions.
\end{acknowledgments}

\bibliography{MQ_QEC}
\clearpage
\appendix
\onecolumngrid

\section{Multiqubit gate construction}
\label{app:MQ_Magnus}
In this section, we derive analytical expressions for $\alpha_j^{(n)}(t)$ and $\varphi_{nm}^{(j)}(t)$, which are used in the analytical expressions and numerical calculations appearing in the main text. We begin from Eq. \eqref{eq:lsf_H}, i.e., the model Hamiltonian in the LD regime. The interaction Hamiltonian in the rotating frame with respect to $H_0$ is:
\begin{equation}
H_I(t)=\sum_{n,j=1}^{N}\eta_j^{(n)} \left(\Omega_n(t)e^{i\omega_0t}\sigma_+^{(n)} +h.c.\right) \left( e^{-i\nu_jt}a_j+h.c. \right).
\end{equation}
For  MS-type gates, the Fourier components of $\Omega_n$ are symmetric around $\omega_0$, therefore, it is convenient to represent it by- $\Omega_n(t)=e^{-i\omega_0t}f_n(t)$, with $f(t)$ a real-valued function. The interaction Hamiltonian then becomes:
\begin{equation}
H_I(t)=\sum_{n,j=1}^{N}\eta_j^{(n)}f_n(t) \sigma_x^{(n)} \left( e^{-i\nu_jt}a_j+h.c. \right).
\end{equation}
The Magnus expansion \cite{Magnus1954} provides a systematic way of deriving the time-evolution operator analytically. The time evolution operator has the general form:
\begin{equation}
U(t)=\exp{\left( \sum_{k=1}^\infty \hat{O}_k\right)},
\end{equation}
where in order $k$, the operator $\hat{O}_k$ is given by $k$ nested integrals of $k-1$ nested commutators of $H_I$. The first two orders read:
\begin{equation}
\begin{aligned}
\hat{O}_1 &= -i\int_0^t dt_1 H_I(t_1), \\
\hat{O}_2 &= -\frac{1}{2}\int_0^t dt_1 \int_0^{t_1} dt_2 \left[H_I(t_1),H_I(t_2)\right].
\end{aligned}
\end{equation}
While the expansion is often truncated at the second order as an approximation to an exact solution, here since $\left[\hat{O}_2,H_I\right]=0$ at all times; this truncation is exact.\\
The expressions for the first two orders are given here:
\begin{equation}
\begin{aligned}
\hat{O}_1 &= \sum_j\left( \sum_{n}\eta_j^{(n)}\sigma_x^{(n)}\int_0^t dt_1 f_n(t_1)e^{-i\nu_j t_1} \right) a_j  - h.c. \\
\hat{O}_2 &= i\sum_{j}\sum_{nm}\eta_j^{(n)}\eta_j^{(m)}\sigma_x^{(n)}\sigma_x^{(m)}\int_0^t dt_1 \int_0^{t_1} dt_2 f_n(t_1)f_m(t_2)\sin(\nu_j(t_1-t_2)) \\
\end{aligned}
\end{equation}
All the above operators are commuting, therefore, we can immediately identify $\alpha_{j}^{(n)}(t)$ and $\varphi_{nm}^{\left(j\right)}\left(t\right)$ from Eq.  \ref{Eq:unitary_lsf}:
\begin{equation}
\begin{aligned}
\alpha_{j}^{(n)}(t) &= \eta_j^{(n)} \int_0^t dt_1 f_n(t_1)e^{i\nu_j t_1} \\
\varphi_{nm}^{\left(j\right)}\left(t\right) &= \eta_j^{(n)}\eta_j^{(m)}\int_0^t dt_1 \int_0^{t_1} dt_2 f_n(t_1)f_m(t_2)\sin(\nu_j(t_1-t_2)) \\
\end{aligned}
\end{equation}

\section{Raman scattering parameters}
\label{app:ramman}
In trapped-ion systems, a significant source of noise, appearing in ground-state or metastable qubits, is Raman scattering. Raman transitions are a standard method to implement transition for these types of qubits, using a virtual intermediate level detuned by $\Delta$ from a, typically short-lived, level. Here specifically we consider qubits encoded on the $4S_{\frac{1}{2}}$ manifold of $40\text{Ca}^+$ ions, and coupled through a Raman transition to the $4P_\frac{1}{2}$ manifold with a laser at $\sim400\text{ nm}$. If the $P$ level becomes populated, spontaneous photon emission occurs, which gives rise to Rayleigh or Raman scattering, inducing bit-flips or phase flips. These effects are reliably modeled as single-qubit depolarization. Elastic and inelastic scattering processes differ slightly, but for simplicity we assume uniform depolarization.

The scattering rate can be estimated using the Kramers-Heisenberg formula ~\cite{ozeri2007errors}:
\begin{equation}
\Gamma_{fi} \propto \frac{\Omega_{1} \Omega_{2}}{\Delta^2} \gamma_P \sim \frac{\Omega^2}{\Delta^2} \gamma_P,
\end{equation}
where $\Omega_{1,2}$ are the Rabi frequencies of the two Raman beams used for the transition, which are detuned by $\Delta$ from the excited level $P$. $\gamma_P$ is the  $P$ level lifetime.

A rough estimate using typical experimental parameters gives
\begin{equation}
\Delta \sim 5~\mathrm{THz} \cdot 2\pi, \quad
\gamma_P \sim 20~\mathrm{MHz} \cdot 2\pi, \quad
\eta = 0.1, 
\end{equation}
\begin{equation}
\Omega_R \sim
\frac{\Omega_{1} \Omega_{2}}{\Delta} \sim
100~\mathrm{kHz}, \quad
\tau = \frac{2\pi}{\Omega_R \eta} \sim 1~\mathrm{ms}.
\end{equation}

The probability of such a scattering event during a gate of duration $\tau$ is therefore
\begin{equation}
\Gamma \sim 1~\mathrm{Hz}, \qquad \text{and} \qquad
P = \Gamma \tau \sim 10^{-3}.
\end{equation}

Note that elastic and inelastic scattering processes differ slightly, but for simplicity we assume uniform depolarization, represented by the jump operators $\sigma_{x}, \sigma_{y}, \sigma_{z}$.

\section{Derivation of bit-flip probabilities under scattering noise}
\label{app:bitflip_analyitcal}
We derive an exact expression (to leading order in $\Gamma\tau$ ) for bit-flip probability of each qubit involved in the gate, given a depolarization noise on qubit $k$. As explained in the main text, we are interested in the case in which the gate is diagonal in $x$ and the noise is $\sigma_k^z$ (or $\sigma_k^y$ ) since in this case the error is not commuting with the gate and there is error propagation. Otherwise it trivially commutes  with the gate.
We start from the RHS of Eq. \eqref{eq:LeadingOrder}
\begin{equation}
\tilde{\rho}(\tau)=(1-\Gamma\tau)\tilde{\rho}(0) +\Gamma\int_0^{\tau}\tilde{C}^{\dagger}\tilde{\rho}(0)\tilde{C} \ dt  + \mathcal{O}((\Gamma\tau)^2),  
\end{equation}
where we already made use of the fact that $\tilde{C}^\dagger \tilde{C}=1$ for scattering, such that the anti-commutator is trivial.

Clearly, within a probability of $1-\Gamma\tau$ , there is no depolarization. However, we are interested in the bit-flip probability of each qubit, given that a noise event did happen. We thus focus in the second part, which is proportional to $\Gamma$. The full expression for this term, with noise $\sigma_k^z$, according to Eq. \eqref{eq:c_transform} is given by:
\begin{equation}
    \int_0^{\tau}\tilde{C}^{\dagger}\tilde{\rho}(0)\tilde{C} \ dt   = \int_0^{\tau} 
    D_j\Bigl( 2\sum_j \alpha_j^{(k)}(t) \, \sigma_x^{(k)} \Bigr) e^{2 i \sum_{n\neq k}\varphi_{kn}(t) \sigma_x^{(k)} \sigma_x^n} \sigma_z^{(k)}  \ \tilde{\rho}(0)  \ \sigma_z^{(k)}  D_j\Bigl(-2\sum_j \alpha_j^{(k)}(t) \,\sigma_x^{(k)} \Bigr) e^{-2 i \sum_{n\neq k}\varphi_{kn}(t) \sigma_x^{(k)} \sigma_x^n}  \ dt .
\end{equation}
 Tracing out the phonons, as in Eq. \eqref{eq:phonon_trace}, yields:
\begin{equation}
    \int_0^{\tau}\tilde{C}^{\dagger}\tilde{\rho}(0)\tilde{C} \ dt 
=  \int_0^{\tau}  e^{2 i \sum_{n\neq k}\varphi_{kn}(t) \sigma_x^{(k)} \sigma_x^n} \left[(1-\gamma_k)\, I \, \tilde{\rho}(0) I  
+ \gamma_k \sigma_x^{(k)} \rho_0(0) \sigma_x^{(k)}\right]  e^{-2 i \sum_{n\neq k}\varphi_{kn}(t) \sigma_x^{(k)} \sigma_x^n} 
\end{equation}
where 
\begin{equation}
\gamma_k = \frac{1 - \exp\Big[-2(2 \bar{n}_j + 1) \sum_j |\alpha_j^{(k)}|^2\Big]}{2} .
\end{equation}
To obtain the bit-flip probability of a qubit $n$  ($n\neq k$) , we should trace out all other qubits. Tracing all qubits $n'\neq n,k$  is straightforward. From the cyclic property of the trace, all $n'$ terms vanish when using the initial ground state of  $\sigma_z$, given by $\rho(0)=\tilde{\rho}(0)=\ket{0}\bra{0}^{\otimes N}$. We are left with:
\begin{equation}
    \text{Tr}_{\sigma_{k}}  \int_0^{\tau}  e^{2 i \varphi_{kn}(t) \sigma_x^{(k)} \sigma_x^n} \sigma_z^{(k)} \left[(1-\gamma_k)\, I \, \ket{00}\bra{00}_{nk}I  
+ \gamma_k \sigma_x^{(k)} \ket{00}\bra{00}_{nk} \sigma_x^{(k)}\right] \sigma_z^{(k)}  e^{-2 i \varphi_{kn}(t) \sigma_x^{(k)} \sigma_x^n}.
\end{equation}
We stress that we are using the initial state of the $z$ ground state for simplicity, but this tracing out procedure can be  easily generalized for any pure $\tilde{\rho}(0)$ which is a product state on our $n$ qubits.
Tracing out qubit $k$, one finds,
\begin{equation}
   \text{Tr}_{\sigma_{n'\neq n}}   \int_0^{\tau}\tilde{C}^{\dagger}\tilde{\rho}(0)\tilde{C} \ dt   =   \int_0^{\tau}  dt \  \sin^2 \left( 2\varphi_{nk}(t) \right)  \ket{1}\bra{1}_{n}+ \cos^2 \left( 2\varphi_{nk}(t) \right)   \ket{0}\bra{0}_{n} .
\end{equation}
Therefore we conclude that the bit-flip probability for qubit $n\neq n',k$ is given by:
\begin{equation}
p_n=\int_0^{\tau} \sin^2 \big( 2\varphi_{nk}(t) \big) \, dt \quad (n \neq k  \text{ for: } \sigma_z^{(k)}) \,
\end{equation}

as in Eq. \eqref{Eq:bitflip}. Clearly, the treatment for the jump operator $\sigma_k^y $ is similar, except that we get another flip due to additional $\sigma_k^x$. So the expression above becomes, 
\begin{equation}
p_n=\int_0^{\tau} \cos^2 \big( 2\varphi_{nk}(t) \big) \, dt \quad (n \neq k  \text{ for: } \sigma_y^k) \ .
\end{equation}

Let us now treat the special case of $n=k$. Now tracing out all $n' \neq k$ is not that simple since they are all entangled with qubit $k$. We therefore take a different approach;
Let us calculate the residual  $2\times2$ density matrix of qubit $k$ , in $x$ basis, after  tracing out all other qubits.  We find that: 
\begin{align}
\textbf{Diagonal terms:}\quad
\frac{1}{2^{N-1}}
\sum
\bra{\ldots \pm_k \ldots}
\tilde{C}^{\dagger}\tilde{\rho}(0)C
\ket{\ldots \pm_k \ldots}
&= \frac{1}{2}
\\[1ex]
\textbf{Off-diagonal terms:}\quad
\frac{1}{2^{N-1}}
\sum
\bra{\ldots \pm_k \ldots}
\tilde{C}^{\dagger}\tilde{\rho}(0)C
\ket{\ldots \mp_k \ldots}
&=
\frac{(1-2\gamma_k)}{2}
\prod_{n\neq k}\cos\!\big(4\varphi_{nk}(t)\big)
\end{align}

where as before $\rho(0)=\tilde{\rho}(0)=\ket{0}\bra{0}^N$  , and the ellipsis, $...$, represent tracing out over all states in the $x$ basis for $n\neq k$, while for the state $k$ we fix $\bra\pm, \ket{\pm}$ for the diagonal elements of the reduced density matrix of qubit $k$, and  $\bra\pm, \ket{\mp}$ for the off-diagonal elements. One can easily obtain the expressions  above when using the $x$ basis. Rotating the above matrix to the $z$ basis, in order to obtain information on the bit-flip probabilities yield:
\begin{equation}
    \rho_k (0) = \big(\frac{1}{2}+\frac{(1-2\gamma_k)}{2}
\prod_{n\neq k}\cos\!\big(4\varphi_{nk}(t)\big)\big) \ket{0}\bra{0} +\big(\frac{1}{2}-\frac{(1-2\gamma_k)}{2}
\prod_{n\neq k}\cos\!\big(4\varphi_{nk}(t)\big)\big)\ket{1}\bra{1}.
\end{equation}
Thus, the bit-flip probability of qubit $k$ is:
\begin{equation}
 p_k= \int_0^\tau   \big(\frac{1}{2}-\frac{(1-2\gamma_k)}{2}
\prod_{n\neq k}\cos\!\big(4\varphi_{nk}(t)\big)\big) dt
\end{equation}

\section{Comparison to MQ full Hilbert space calculation and sequential two-qubit protocols   }
\label{app:analytical_MS}
\subsection{Reproducing previous result for MS gate}
We present two examples demonstrating that our analytical approach to qubit depolarization or dephasing significantly reduces the numerical overhead, which is otherwise required when solving the full Hilbert space of both spins and phonons via an exact Lindbladian simulation of the density matrix.

 In Ref.~\cite{Schwerdt2022} an exact dynamics simulation was performed to obtain the bit-flip distribution in the case of  an MS gate. However, this distribution can be calculated directly using Eq.~\eqref{eq:p_nvec}. The probability that a scattering event occurred but none of the connected qubits flipped is obtained by retaining only the $\cos^2\!\left(2\varphi_{nk}(t)\right)$ terms in Eq.~\eqref{eq:p_nvec}. If we further require that the faulty qubit itself has not flipped, the $(1-\gamma_k)$ term in Eq.~\eqref{eq:phonon_trace} should be included.
For the MS gate, we use only the COM mode. The expressions for $\alpha(t),\varphi(t)$ in this case are well known and are given by (~\cite{Srensen1999,Srensen2000}):
 \begin{equation}
     |\alpha^{(n)}_j(t)|^2=\sin^2 \big(\frac{\xi t}{2} \big),\ \varphi_{nm}(t)=\frac{\xi t -\sin\bigl(\xi t \big)}{8} 
 \end{equation} 
 and the integral for a 5 qubit register is:
 \begin{equation}
     p_{\text{no-flips}}=\Gamma\int_0^\tau (1-\gamma_k)\cos^8(2\varphi(t)) dt =
\Gamma\tau\int_{0}^{2\pi}
\frac{\left(1 + e^{-2\sin^2(u/2)}\right)\,
\cos^{8}\!\left(\frac{u}{4} - \frac{\sin u}{4}\right)}{4\pi}
\, du
= 0.291391
 \end{equation} 
Where we have substituted the expressions for $\alpha$ and $\varphi$, followed by a simple change of variables $u=\xi t$.
Note that $ p_{\text{no-flips}}$ with $N=5$, is equivalent to $\mathbf{n}=(0,0,0,0,0)$ in \cref{eq:p_nvec}.

By switching integration limits we find that:
  \begin{equation}
     p_{\text{no-flips}}=
\Gamma\tau\int_{0}^{2\pi}
\frac{\left(1 + e^{-2\sin^2(u/2)}\right)\,
\cos^{8}\!\left(\frac{u}{4} - \frac{\sin u}{4}\right)}{4\pi}
\, du
= \Gamma\tau\int_{0}^{2\pi}
\frac{\left(1 + e^{-2\sin^2(u/2)}\right)\,
\sin^{8}\!\left(\frac{u}{4} - \frac{\sin u}{4}\right)}{4\pi}
\, du =p_{\text{all-flip/{faulty}}}
 \end{equation}
That is, the probability of all connected qubits are flipped except for the faulty qubit ($\mathbf{n}=(1,1,1,1,0)$) is equal to the probability all qubits were not flipped. Similarly the probability that all qubits are flipped ($\mathbf{n}=(1,1,1,1,1)$) is given by,
  \begin{equation}
     p_{\text{all-flipped}}
= \Gamma\tau\int_{0}^{2\pi}
\frac{\left(1 - e^{-2\sin^2(u/2)}\right)\,
\sin^{8}\!\left(\frac{u}{4} - \frac{\sin u}{4}\right)}{4\pi}
\, du = 0.081.
\label{Eq:app_ms_all_fliped}
\end{equation}
We can use again the same trick of integration limits and deduce that $p_{\text{no-flip}/{faulty}}=0.081$. These results cover with few simple integrals the reported probabilities of the first and last rows in Table I of Ref.~\cite{Schwerdt2022}, within the numerical precision. This analytical approach is not only efficient, but also allow to asses these probabilities in the multi-qubit, multi-mode and multi-tone gate, where a full numerical of the entire Hilbert space is out of reach.

\subsection{Distribution of Hook Errors Under MS Evolution}

Assuming MS evolution, with the above $\varphi_{nm}(t)$
and tracing out one of the qubits, defined as the faulty qubit, we can obtain the distribution for the number of propagated spin flips shown in \cref{fig:scattering}\textcolor{blue}{c}.

The probability to observe $n$ flipped qubits is given by

\[
 p(n)=\frac{1}{2\pi}\int_0^{2\pi}du\,
\binom{N}{n}
\sin^{2n}\!\bigl(\varphi(u)\bigr)
\cos^{2N-2n}\!\bigl(\varphi(u)\bigr).
\]

For $N=5$, the resulting distribution is shown (dashed black) in Fig.~1(c) . 
In the large $N$ limit, the distribution exhibits distinct scaling behavior at the edges ($n=0$ or $n=N$) and in the bulk ($n\simeq N/2$). Near the edges, 
one finds (by expanding around the saddle point $u=2\pi$) that for $N\gg1$
\[
 p(0), p(N)
 \approx\frac{3^{1/3}\Gamma\!\left(\frac16\right)}
{6\pi}
N^{-1/6}
\approx
0.43N^{-\frac{1}{6}}
\]

At the center of the distribution (by expanding around the saddle point $u=\pi$), we find that
\[
 p\!\left(\frac{N}{2}\right) \approx \frac{1}{\pi N}.
\]

These behavior should be contrasted with the standard hook errors for discrete two-qubit gates, in which an error on the ancilla qubit can propagate uniformly to any subset of the data qubits. In that case, the probability distribution is just
\[
p(n)\approx \frac{1}{N}.
\]

The collective phase accumulation inherent to the MS gate therefore suppresses harmful hook errors. In particular, the probability for the worst-case event, in which approximately half of the data qubits experience correlated errors due to propagation from the ancilla near the middle of the gate, is reduced by a factor of order $\pi$ compared to the standard two-qubit construction. 

However, this suppression is not parametric. In an MQ gate, any subset of approximately $n=N/2$ qubits may experience correlated errors, unlike the case of a specifically ordered sequence of two-qubit gates, where the propagation structure is more constrained. This combinatorial enhancement compensates for the reduction in the probability associated with a given error pattern, preventing a parametric advantage over the standard two-qubit implementation.

\begin{figure}
\label{fig:largeN_hool}
    \centering
    \includegraphics[width=0.5\textwidth]{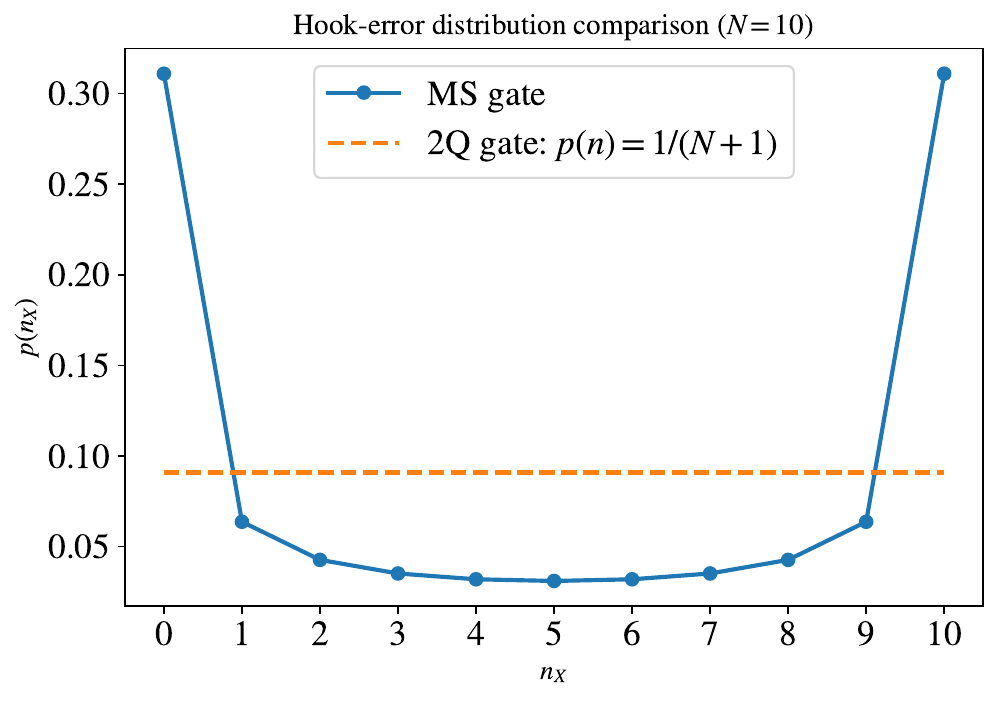}
    \caption{Comparison between the probability distributions for the number of data qubits flipped due to hook errors, contrasting a sequence of discrete two-qubit gates with an MQ gate. Here, $N=10$ data qubits are considered, together with an additional ancilla qubit that was traced out. }
    \label{fig:2q_MQ_compare}
\end{figure}

\cref{fig:2q_MQ_compare} shows the comparison between MQ and two-qubit gate for the case of $N=10$ data qubits (corresponding to a total of $11$ qubits including the ancilla, which is subsequently traced out).

These results suggest that multi-qubit (MQ) gates may nevertheless provide some advantage for implementing heavier stabilizers while partially mitigating the impact of hook errors.

\section{Multiqubit gate dynamics}
\label{app:gate_param}
\subsection{gate parameters}
Here we provide additional details regarding the MQ gate used to obtain the results in Fig.~\ref{fig:scattering}. Our gate design protocol follows Ref.~\cite{shapira2023fastdesignscalingmultiqubit}. The gate couples to the radial (transverse) modes of an equidistant crystal of trapped $^{40}\mathrm{Ca}^+$ ions.
To implement the desired unitary in Eq.~\eqref{Eq:unitary_lsf}, we must satisfy $2N$ linear constraints for each ion:
\begin{equation}
\mathrm{Re}\!\left(\sum_j \alpha_j^{(n)}\right)=0, \quad
\mathrm{Im}\!\left(\sum_j \alpha_j^{(n)}\right)=0,
\end{equation}
as well as $N(N-1)/2$ entangling phases $\varphi_{nm}$. This implies that each ion requires on the order of $N$ degrees of freedom, which we realize using distinct tone amplitudes. In total, the construction involves $\sim N^2$ degrees of freedom. 
To provide sufficient flexibility for solving the optimization problem, we choose approximately $8$ tones between any two neighboring modes, corresponding to $8N$ degrees of freedom per ion. For the $d=5$ rotated surface code with $N=37$ qubits (data and ancilla qubits), this results in a total of $297$ tones. The resulting gate time is $\tau \approx 321~\mu\mathrm{s}$.

\begin{figure}
    \centering
\includegraphics[width=0.9\textwidth]{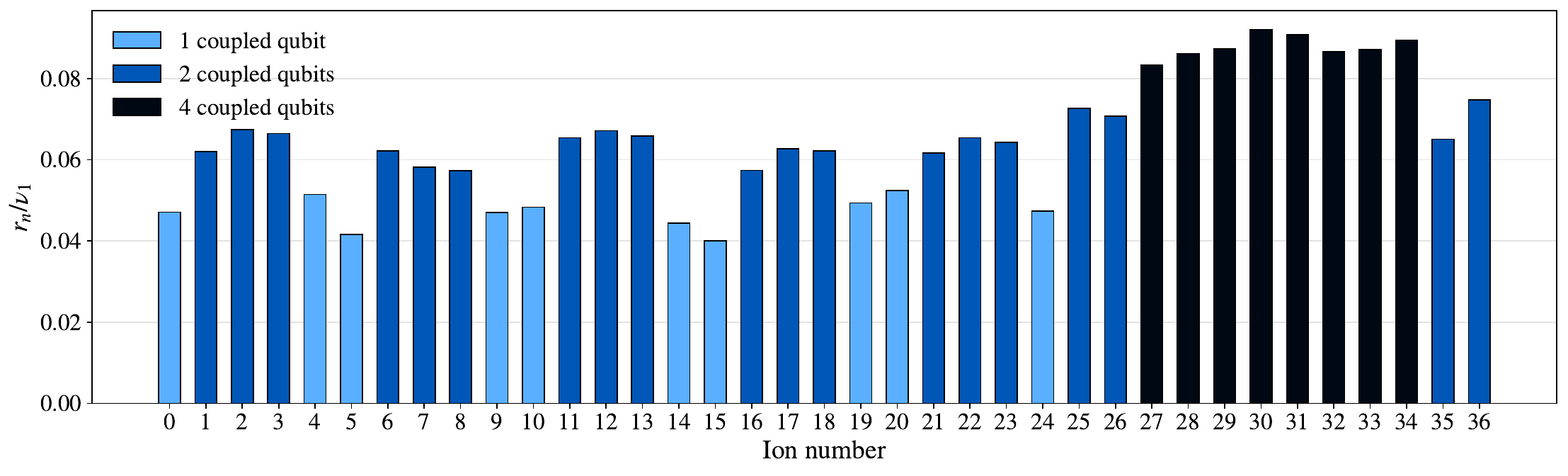}
    \caption{Total Rabi drive on each ion, $r_n=\sqrt{\sum_i |\Omega_{n,i}|^2}$, where $i$ denotes the tone index and $n$ the ion index, normalized by the lowest mode frequency, $\nu_1=3.04~\mathrm{MHz}$. Different colors indicate the number of coupled qubits, clearly demonstrating that the Rabi drive increases with the number of couplings.}
    \label{fig:tones}
\end{figure}

In Fig.~\ref{fig:tones}, we present the total Rabi drive on each ion \(n\) (for the gate shown in \cref{fig:scattering}), defined as
\[
r_n=\sqrt{\sum_i |\Omega_{n,i}|^2},
\]
where \(i\) labels the different tones. The results are normalized with respect to the lowest motional mode frequency, \(\nu_1 = 3.04~\mathrm{MHz}\).

Averaging over all ions, we obtain an effective Rabi frequency of
\[
\bar{\Omega}_R \sim 200~\mathrm{kHz}.
\]
Using the analysis presented in App.~\ref{app:ramman}, with a detuning of \(\Delta =2\pi \cdot 5~\mathrm{THz}\), we estimate
\begin{equation}
\Gamma_s \tau
=
\frac{\bar\Omega_R \gamma_P}{\Delta}\tau
\approx
\left(\frac{1}{5 \cdot 2\pi\,\mathrm{THz}}\right)
\left(20 \cdot 2\pi\,\mathrm{MHz}\right)
\left(321\,\mu\mathrm{s}\cdot 200\,\mathrm{kHz}\right)
=
2.568 \times 10^{-4}.
\end{equation}
This confirms that the first-order expansion of the Lindbladian remains well justified.

\subsection{Additional results on error propagation under scattering noise }
\begin{figure}

    \centering
    \includegraphics[width=0.68\textwidth]{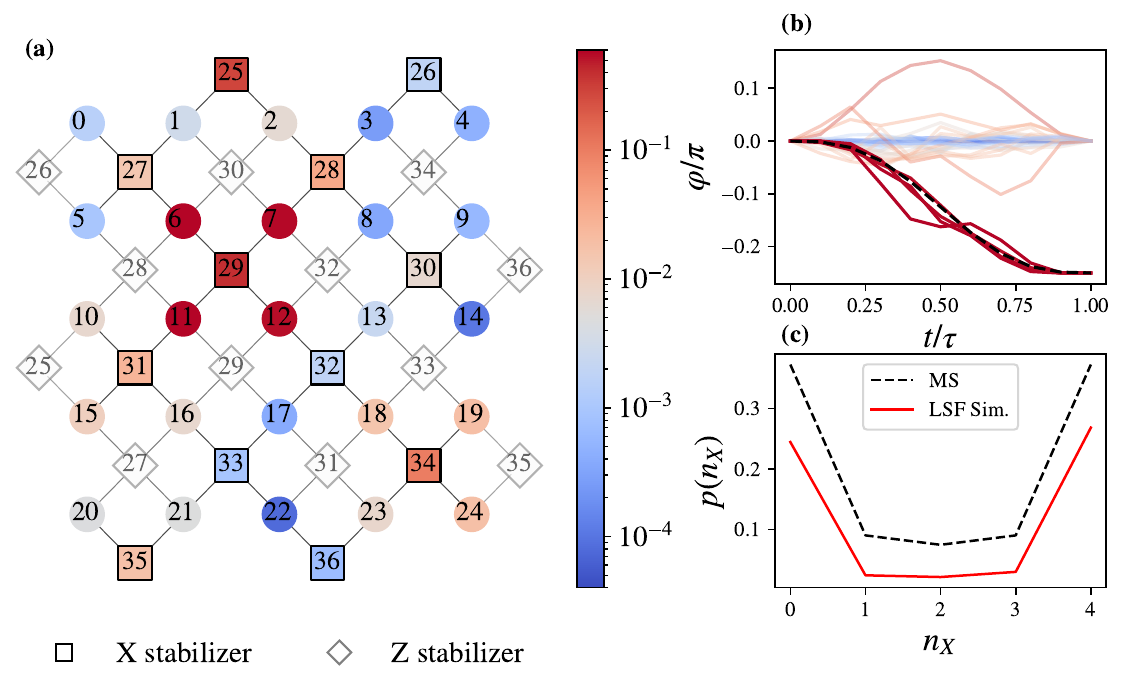}
    \caption{A case study of undesired gate design which leads to large deviation of disconnected $\varphi_{nk}$  }
\label{fig:bad_scatter}
\end{figure}

As discussed in Sec.~\ref{section:analytical}, error propagation  strongly depends on the trajectory $\varphi_{nk}(t)$ where $\{n,k\}$ are disconnected at gate time ($\varphi_{nk}(t=\tau)=0$). In Fig.~\ref{fig:bad_scatter} we present an example of undesired error propagation arising from a large deviation of $\varphi_{nk}(t)$ from zero during the trajectory. Note that in Fig.~\ref{fig:bad_scatter}\textcolor{blue}{c} a significant part of the probability mass is included in long range correlated errors (the red line, representing the correlation between connected qubits, covers only $\approx60\%$  probability mass- where in Fig.\ref{fig:scattering}\textcolor{blue}{c} it covers $95\%$).

By comparison with the favorable case shown in Fig.~\ref{fig:scattering}, it becomes clear that the trajectories of $\varphi_{nk}(t)$ serve as a useful indicator of the level of error propagation to disconnected qubits.

\begin{figure}

    \centering  \includegraphics[width=0.68\textwidth]{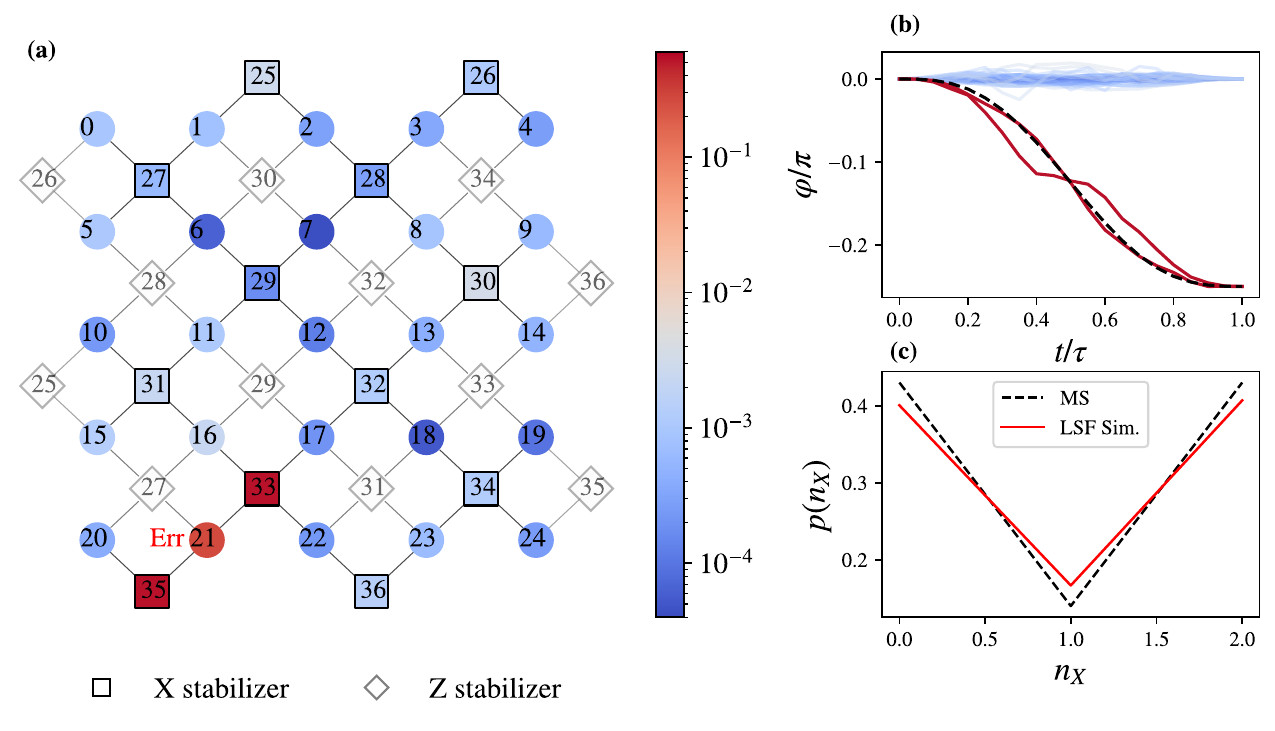}
    \caption{The gate used in Fig.\ref{fig:scattering} 
    but with an error assigned to the data qubit $k=21$ }
\label{fig:app_scatter_data}
\end{figure}

We also examine in Fig.\ref{fig:app_scatter_data} the case in which the faulty qubit is a data qubit, $k=21$. In this case the connected qubits are ancilla qubits. 
\\

\section{Derivation for of the Pauli channel matrix elements $\eta_{ij}$}
\label{app:eta_ij}
We start from Eq.~\eqref{eq:x_channel},
\begin{equation}
\tilde\rho(\tau) = \sum_{ij} \eta_{ij}(\tau)\, P_i\, \tilde\rho(0)\, P_j.
\end{equation}
where the $P_i$s are Pauli strings of either $\sigma_x$ or $I$ operators, according to the binary vector, $i$. Starting with the LHS, by solving the ODE in Eq.~\eqref{eq:heating_rho_ode}
we can obtain the $X$ basis representation of $\rho(t)$. We write it as:
\begin{equation}
\rho(t) = \sum_{i,j} \rho^{(X)}_{ij}(t)\, |{\sigma}_x^i\rangle\langle {\sigma}_x^j|
\end{equation}

We want the matrix elements of $\rho(t)$ in the $Z$ basis.
By definition,
\begin{equation}
\rho^{(Z)}_{k\ell}(t)
= \langle {\sigma}_z^k | \rho(t) | {\sigma}_z^\ell \rangle .
\end{equation}

Substituting the expansion in the $X$ basis,
\begin{equation}
\rho^{(Z)}_{k\ell}(t)
= \sum_{i,j} \rho^{(X)}_{ij}(t)
\langle {\sigma}_z^k | {\sigma}_x^i \rangle
\langle {\sigma}_x^j | {\sigma}_z^\ell \rangle .
\end{equation}

Using the overlaps,
\begin{equation}
\langle {\sigma}_z^k | {\sigma}_x^i \rangle = H_{k i},
\qquad
\langle {\sigma}_x^j | {\sigma}_z^\ell \rangle = H^\dagger_{j \ell},
\end{equation}
with $H_{ki}$ the $k,i$ element of the $n$-qubit Hadamard transform, $H^{\otimes n}$. We obtain,
\begin{equation}
\rho^{(Z)}_{k\ell}(t)
= \sum_{i,j} H_{k i}\, \rho^{(X)}_{ij}(t)\, H^\dagger_{j \ell}.
\end{equation}

In matrix form,
\begin{equation}
\rho^{(Z)}(t) = H\, \rho^{(X)}(t)\, H^\dagger .
\end{equation}
with $H$ given in $Z$ basis (or $X$ which is the same).

Now for the RHS,
\begin{equation}
\rho^{(Z)}_{k\ell}=\langle {\sigma}_z^k | \sum_{i,j} \eta_{i j}\,
{\sigma}_x^i \, \rho_0 \, {\sigma}_x^j \, | {\sigma}_z^\ell \rangle
\end{equation}

We take the initial state to be 
\begin{equation}
\rho_0 = |0_z\rangle\langle 0_z|^{\otimes N},
\end{equation}
We insert it explicitly:
\begin{equation}
\rho^{(Z)}_{k\ell}= \sum_{i,j} \eta_{i j}\,
\langle {\sigma}_z^k | {\sigma}_x^i | 0 \rangle
\langle 0 | {\sigma}_x^j | {\sigma}_z^\ell \rangle .
\end{equation}
clearly if $i,j,k,\ell$ are bitstrings:
\begin{equation}
\langle {\sigma}_z^k | {\sigma}_x^i | 0 \rangle = \delta_{k i},
\qquad
\langle 0 | {\sigma}_x^j | {\sigma}_z^\ell \rangle = \delta_{j \ell},
\end{equation}
we obtain
\begin{equation}
\rho^{(Z)}_{k\ell}= \sum_{i,j} \eta_{i j}\, \delta_{k i}\, \delta_{j \ell}
= \eta_{k \ell}.
\end{equation}

Therefore,
\begin{equation}
\rho^{(Z)}_{k\ell}(t)
= \bigl( H\, \rho^{(X)}(t)\, H^{T} \bigr)_{k\ell}
= \eta_{k\ell}.
\end{equation}
as required.
\section{Heating and Motional dephasing in Multiqubit gates}
\label{app: heating_LSF}
\subsection{Heating in MQ gates}

In Section~\ref{sec:x_dephasing}, we showed that, for the MS gate, heating events effectively behave as single bit-flip errors. Here, we generalize this treatment to MQ gates. Starting from Eq.~\eqref{eq:heating_rho_ode}, and integrating under the initial condition of the motional ground state in the \(z\)-basis, we obtain
\begin{equation}
    \label{eq:app_heating_rxx}
    \rho_{x,x'}(\tau)
    =
    \frac{1}{2^N}
    e^{-\sum_{n,m=1}^{N}A_{nm}(\tau)(x_n-x_n')(x_m-x_m')},
\end{equation}
where \(x,x'\) are binary vectors of size \(N\) representing spin configurations in the \(x\)-basis. The coefficients \(A_{nm}\) are given by
\begin{equation}
    A_{nm}(\tau)
    =
    \sum_j
    \Gamma_j
    \frac{2\bar{n}_j+1}{2}
    \int_0^\tau
    \alpha_j^{(n)}(t)\alpha_j^{*(m)}(t)\,dt .
\end{equation}

Using \cref{eq:had}, the bit-flip probabilities are given by
\begin{equation}
    \eta_{z,z}(\tau)
    =
    \sum_{x,x'}
    \frac{(-1)^{z(x+x')}}{4^N}
    e^{-\sum_{n,m=1}^{N}A_{nm}(\tau)(x_n-x_n')(x_m-x_m')}
    =
    \mathcal{O}(\Gamma^{|z|}),
\end{equation}
where \(|z|\) denotes the Hamming weight of \(z\) relative to the computational ground state, and \(zx\), \(zx'\) denote binary-vector dot products.

It is straightforward to verify that whenever \(z_n=1\), the zeroth-order contribution vanishes since
\[
\sum_{x_n,x_n'=\pm1}(-1)^{x_n+x_n'}=0.
\]
The same cancellation occurs for every additional flipped qubit \(z_{n'}=1\), implying that each spin flip contributes at least one power of \(\Gamma\). In general, there is no reason for the linear term itself to vanish, and therefore the dominant contribution scales exponentially with the number of flips.

\begin{figure}
    \centering
    \includegraphics[width=0.5\textwidth]{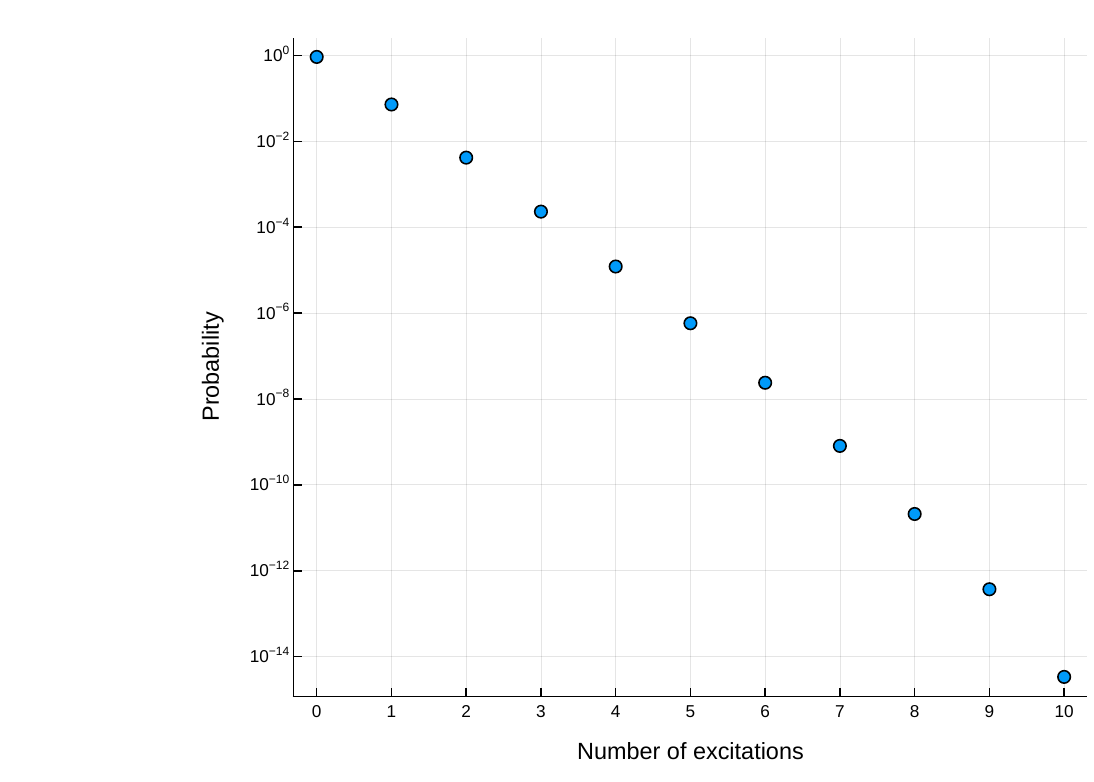}
    \caption{
    An example of an MQ gate with \(N=10\) implementing all-to-all connectivity, similarly to the MS gate. This gate utilizes all the motional modes to generate the all-to-all interaction, and the heating rate is a constant 0.1 quanta per gate time for all motional modes. We do not use robustness constraints to mitigate the effect of heating for this gate. We observe dominant single-bit-flip behavior, analogous to the MS case shown in \cref{fig:n6}\textcolor{blue}{a}. 
}
    \label{fig:app_heating_MQ}
\end{figure}

We can explicitly evaluate the integrals entering \(A_{nm}\) in order to obtain the reduced spin density matrix after tracing out the phonon degrees of freedom. Numerically, we observe the same linear scaling behavior on a logarithmic scale for all-to-all connectivity, now involving many modes and tones. An example of heating in an MQ gate is presented in \cref{fig:app_heating_MQ}. This confirms that, to first order, the heating-induced error channel in MQ gates is well approximated by a product of independent single-qubit bit-flip channels.

\subsection{Motional dephasing in MQ gates }

In this section, we compare pairs of qubits that are coupled,
$\varphi_{nm}(\tau)=\pi/4$, with uncoupled pairs for which
$\varphi_{nm}(\tau)=0$. In particular, we demonstrate how the requirement of generating entanglement between ions $n,m$ gives rise to stronger correlations in the motional-mode displacements.

To quantify this effect, we analyze the correlations both as a function of time and as a function of the motional modes. Using Eq.~\ref{eq:dephasing_2q}, we independently perform either the summation over the mode index $j$ or the integration over time, and subsequently average over all ion pairs satisfying either
$\varphi_{nm}(\tau)=0$ or $\varphi_{nm}(\tau)=\pi/4$.

\begin{figure}[t]
    \centering
    \includegraphics[width=0.8\textwidth]{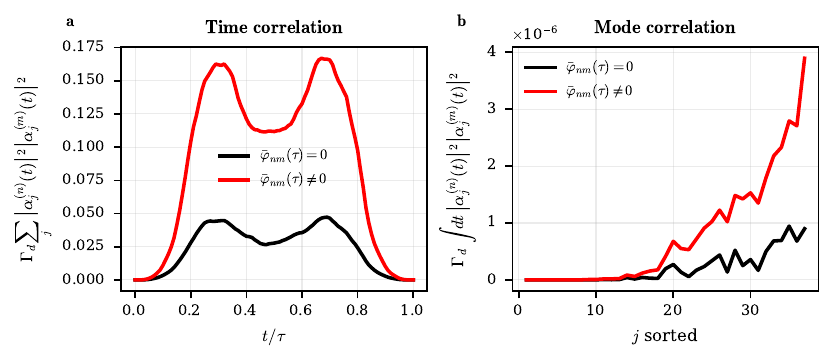}
    \caption{
Comparison of the correlations of coupled and uncoupled qubits as a function of time and mode index. The red curve represents the average over all coupled qubits, while the black curve represents the average over all uncoupled qubits. In the mode-correlation panel, the modes are ordered according to the difference between the two averaged curves to improve visualization clarity.
    }
    \label{fig:corr}
\end{figure}

The resulting correlations are shown in \cref{fig:corr}. We observe a substantially stronger correlation for ion pairs with nonzero entangling phase, $\varphi_{nm}(\tau)=\pi/4$, compared to uncoupled pairs. This behavior highlights the intrinsic connection between entangling interactions and correlated motional displacements in MQ gates.

\section{QEC Simulation Framework}
\label{app:qec_details}
In this section, we provide additional details regarding our QEC simulations. The surface code grid was constructed in a manner similar to the illustration in Fig.~\ref{fig:scattering}, with $d^2$ data qubits and $(d^2 - 1)/2$ ancilla qubits, where the same ancilla qubits are used for both $X$ and $Z$ stabilizers.
Noise was inserted before each MQ gate. As explained in the text, the noise model was constructed by separating the noisy process occurring prior to the gate from the subsequent ideal evolution. 
Since we use Stim to run our simulations, which is naturally built for two-qubit gates, we implement the stabilizer measurements using the standard CNOT-based circuit structure.
We neglect the additional errors associated with the residual single-qubit operations applied before and after the MQ \(R_{xx}\) and \(R_{zz}\) gates in order to make them equivalent to CNOT gates. Since our model already includes single-qubit error channels arising from heating, scattering, and idle errors, we do not expect these additional contributions to qualitatively modify the overall picture. The same argument holds for measurement error.
Of course, such contributions should be included in a more applied and detailed benchmarking analysis. Under this assumption, the ideal part of the MQ gate can be equivalently represented by CNOT gates, and all noisy events (Pauli strings) are applied prior to any entangling operation.

The error channel due to scattering is based on the assumption that the $\varphi_{nm}(t)$ follow the MS evolution, see Eq. \eqref{eq:ms_phi}. Therefore we can build a table of probabilities of all possible correlated errors, similar to Table I in Ref.~\cite{Schwerdt2022}. An error propagation will occur only if the Pauli error anti-commutes with the gate. Otherwise only the faulty qubit experience an error.
For example, the noise channel when measuring $X$ stabilizer, for the faulty qubit is $k$, is given by:
\begin{align}
\mathcal{\epsilon}_{k}^{(X)}(\rho)
&= (1 - p)\, I \rho I
+ \frac{p}{3}\, \sigma_x^{(k)} \rho \sigma_x^{(k)} \\
&\quad + \frac{p}{3}\left[
\sigma_z^{(k)}
\left(
\sum_{\mathbf{i}} q_{\mathbf{i}} \, P_{\mathbf{i}}\, \rho \, P_{\mathbf{i}}
\right)
\sigma_z^{(k)}
+
\sigma^y_k
\left(
\sum_{\mathbf{i}} q_{i} \, P_{\mathbf{i}} \, \rho \, P_{\mathbf{i}}
\right)
\sigma^y_k
\right].
\end{align}

Here, $P_\mathbf{i}$ labels a Pauli string. Each Pauli string consists of either $\sigma_x^{(j)}$ or $I^{(j)}$ acting on every qubit $j$ that is coupled to the faulty qubit $k$. The probability $q_{\mathbf{i}}$ associated with a given flip event, represented by the Pauli string $\mathbf{i}$, can be derived from Eqs.~\eqref{eq:phonon_trace} and \eqref{eq:p_nvec}  (see App.~\ref{app:analytical_MS}). 
When measuring $Z$ stabilizers, the noise channel is obtained by exchanging $X \leftrightarrow Z$ in the above definition.

\begin{figure}
    \centering    
    \includegraphics[width=.98\textwidth]{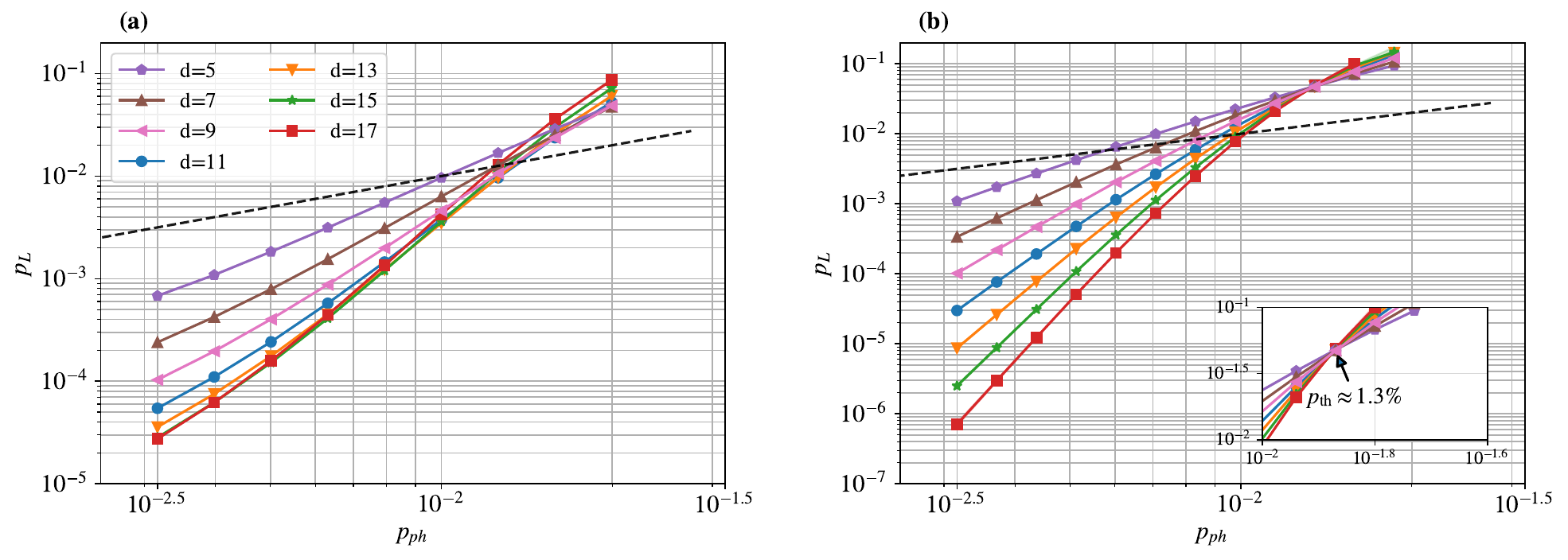}
    \caption{
    Same comparison as in \cref{fig:surface_code_qec}, but for larger code distances.
    }
    \label{fig:qec_large_d}
\end{figure}

Lastly, in \cref{fig:qec_large_d} we present the same comparison as in \cref{fig:surface_code_qec}, but for larger code distances. As before, \textcolor{blue}{\cref{fig:qec_large_d}a} corresponds to the case where all qubit pairs are assigned the same two-qubit error rate, $p_{2q}=3\times10^{-5}$, while \textcolor{blue}{\cref{fig:qec_large_d}b} considers two-qubit errors only between directly coupled qubits.

Indeed, similarly to Ref.~\cite{Fowler2014Nonlocal}, we find that in \cref{fig:qec_large_d}\textcolor{blue}{a} the improvement in the code performance saturates, such that there is essentially no further gain beyond $d\approx17$. In contrast, in \cref{fig:qec_large_d}\textcolor{blue}{b} we observe a clear threshold behavior, since only coupled qubits can experience two-qubit errors, making the total error rate per qubit intensive.

\end{document}